\documentclass[12pt]{article}
\usepackage{latexsym}
\usepackage{fontenc}
\usepackage{amssymb}
\usepackage{amsmath}
\usepackage[all]{xy}
\usepackage[dvipdf]{epsfig}
\usepackage{color}
\usepackage{graphicx}
\usepackage{amsfonts}
\newcommand{\EA}[1]{\Gamma_{#1}}
\newcommand{\BA}[1]{S_{#1}}
\newcommand{\vol}{\int\!\!\textrm{d}^dx\,}
\numberwithin{equation}{section}

\begin{document}
\title{\begin{flushright}
\normalsize{MZ-TH/08-32}
\bigskip
\end{flushright}
Bare Action and Regularized Functional Integral of Asymptotically Safe Quantum Gravity}
\date{}
\author{Elisa Manrique 
and Martin Reuter\\
Institute of Physics, University of Mainz\\
 Staudingerweg 7, D-55099 Mainz, Germany}
\maketitle

\begin{abstract}
Investigations of Quantum Einstein Gravity (QEG) based upon the effective average action employ a flow 
equation which does not contain any ultraviolet (UV) regulator. Its renormalization group trajectories emanating
from a non-Gaussian fixed point define asymptotically safe quantum field theories. A priori these theories 
are, somewhat unusually, given in terms of their effective rather than bare action.  In this paper we
construct a functional integral representation of these theories. We fix a regularized measure and show that every trajectory of effective average actions, depending on an IR cutoff only, induces an associated trajectory
of bare actions which depend on a UV cutoff. Together with the regularized measure these bare actions give
rise to a functional integral which reproduces the prescribed effective action when  the UV cutoff is removed.
In this way we are able to reconstruct the underlying microscopic (``classical'') system and identify its fundamental degrees
of freedom and interactions.
The bare action of the Einstein-Hilbert truncation is computed and its flow is analyzed as an example.
Various conceptual issues related to the completion of the asymptotic safety program are discussed. 
\end{abstract}


\section{Introduction}\label{introduction}

The problem of finding a fundamental quantum theory of  gravity is still an exciting challenge which is pursued within a 
variety of approaches \cite{kiefer,A,R,T}.
In the context of the  asymptotic safety program \cite{wein}-\cite{livrev}, for instance,
a lot of efforts were devoted to establishing the existence of an ultraviolet fixed point at which Quantum Einstein
 Gravity (QEG) can be renormalized. Detailed calculations revealed that the renormalization group (RG) flow of 
the theory  does indeed possess an appropriate non-Gaussian fixed point (NGFP) in all approximations which were investigated.

Formulated in terms of the gravitational average action  as proposed in \cite{mr},
the RG flow in question is that of the {\it{effective average action}} $\EA{k}[g_{\mu\nu},\cdots]$, henceforth abbreviated EAA \cite{avact},\cite{ymrev}.
While similar in spirit to the idea of a Wilson-Kadanoff renormalization, it replaces the iterated coarse graining 
procedure 
by a direct mode cutoff at the infrared (IR) scale $k$. More importantly,
the EAA is a scale dependent version of the ordinary { \it{effective}} action, while a ``genuine'' Wilsonian action 
$\BA{\Lambda}^{W}$ is a {\it{bare}} action, i.e. it is to be used under a regularized path integral.
As a result, it depends on
 the ultraviolet (UV) cutoff $\Lambda$; its  dependence on $\Lambda$ 
 is governed by a RG equation which is different from that for $\EA{k}$.

In a sense, $\BA{\Lambda}^{W}$ for different values of $\Lambda$ is a set of actions for the same system: 
the Green's functions have to be computed from $\BA{\Lambda}^{W}$ by a 
further functional integration over the low momentum modes, and this integration renders them  independent of $\Lambda$. 
By contrast, the EAA can be thought of as the standard effective action for a family of different systems: for any
 value of $k$ it equals the standard effective action of a model with the bare action $\BA{\Lambda}+ \Delta_{k}S$
 where $\Delta_{k}S$ is the mode suppression term. The corresponding $n$-point functions do depend on $k$;  
they provide an effective field theory description \cite{bh}-\cite{mof}
of the physics at scale $k$. These $n$-point functions are simply the functional derivatives of $\EA{k}$, 
so their computation requires no further functional integration. (See \cite{avactrev,livrev} for a disscussion of this point.)

In the EAA framework, a quantum field theory  is fully defined once a {\it{complete}} RG trajectory has been
 constructed, that is, a solution of the functional RG equation (FRGE) for $\EA{k}$ which is well defined for 
all $k\in [0,\infty)$. In particular it must be free from divergences in the IR ($k\rightarrow 0 $) and the UV 
($k\rightarrow\infty$). In asymptotically safe theories the latter condition is met by arranging the RG trajectory
 to hit the NGFP in the limit $k\rightarrow\infty$. Given a complete $\EA{k}$ trajectory we have, in principle, 
complete knowledge of all properties of the quantum theory at hand. Its Green's functions are the derivatives of 
 $\EA{k}$ and at $k=0$ they  coincide with those of the standard effective action $\EA{}\equiv\EA{k=0}$ \cite{avact}.

Because of these differences between the EAA and a genuine Wilson action, this way of constructing 
an asymptotically safe field theory does not by itself yield  a regularized path integral over metrics whose
``continuum limit" would be related to the  RG trajectory $\{\Gamma_k, 0\leq k<\infty\}$ in a straightforward way. 
In order to understand
 this important point let us recall how the EAA was employed in gravity up to now.

The starting point for the definition of the EAA and the derivation of its flow equation is a path integral over metrics,
$\int\mathcal{D}\gamma_{\mu\nu}\exp{(-S[\gamma_{\mu\nu}])}$ which is UV regularized in some way. This path integral
is reformulated in a background field language, gauge fixed, augmented by source terms, and then a mode suppression term
$\Delta_k S$ is added to $S$. By definition, the EAA is essentially the Legendre transform of the resulting generating 
functional. Its FRGE is found by straightforwardly applying the scale derivative $\partial_k$ to this definition.

One of the salient features of this FRGE is that it continues to be well behaved in the ultraviolet {\it{even when the 
UV regulator originally built into the path integral is removed}}\footnote{This requires an appropriate fall-off behavior of
the coarse graining kernel, which we always assume in the following.}. Roughly speaking the reason is that 
$k\partial_k\EA{k}$ receives  contributions only from modes with covariant momenta near (or below) $k$. 
In fact, in all investigations which employed the gravitational EAA so far\cite{mr}-\cite{je1}, the FRGE without an UV regulator 
(``$\Lambda$-free FRGE") has been used. For instance, the NGFP  that has been discovered refers to the 
RG flow implied this ``$\Lambda$-free" equation.

The UV regularization being superfluous at the FRGE level has both advantages and disadvantages. Clearly, the major 
advantage is that it allows us to search for asymptotically safe theories directly at the effective level, without the
additional burden of having to  construct a regularized path integral and control its infinite cutoff limit.

Note that the question of whether or not a theory is asymptotically safe is decided by the properties of the effective - as 
opposed to the bare- action since the former is directly related to S-matrix elements, say. They are free from 
 divergences if their generating functional $\Gamma$ is so, and this in turn is the case when 
$\EA{}\equiv\EA{k=0}$ is connected to a fixed point $\EA{*}$ by a complete, everywhere regular RG trajectory
$\{\EA{k}, 0\leq k<\infty \}$. On the other hand, the relation between the $S$-matrix elements and the bare action $\BA{\Lambda}$ that would enter
a regularized path integral $\int\mathcal{D}_{\Lambda}\gamma_{\mu\nu}\exp{(-S_{\Lambda}[\gamma_{\mu\nu}])}$ is much more indirect. A priori we do not know which behavior of $\BA{\Lambda}$ for $\Lambda\rightarrow \infty$
corresponds to the absence of  divergences in observable quantities 
and to a fixed point $\EA{*}=\lim_{k\rightarrow\infty}\EA{k}$. In general 
the relationship between $\EA{k}$ and $\BA{\Lambda}$ will depend on how we regularize the path integral measure 
$\mathcal{D}_{\Lambda}\gamma_{\mu\nu}$.

Working with the $\Lambda$-free FRGE, the advantage is that all problems related to the path integral, its bare action and 
measure, can be sidestepped. The corresponding disadvantage is that even if we knew an exact, complete RG trajectory 
$\{\EA{k}, 0\leq k<\infty\}$ which would amount to a well defined quantum field theory, we would not have a path 
integral formulation of this theory at our disposal, and we could not even be sure that such a formulation actually exists.

Conceptually there is nothing wrong with that. For systems with finitely many degrees of freedom  canonical quantization, 
path integral quantization and the quantization by a FRGE are equivalent but we cannot be sure that this equivalence will 
always hold true in quantum field theory. In principle the $\Lambda$-free FRGE could yield a physically completely 
satisfactory quantum field theory, predictive and consistent, but has no path integral representation. 
However, in the following we shall
argue that this is actually not what happens in QEG.

In fact, in this paper we are going to demonstrate that it is possible to ``reconstruct" a regularized functional  
integral in such a way that it describes a fixed, prescribed asymptotically safe theory in the infinite cutoff limit.
This path integral representation is not ``canonically" given, however, as it requires an extra ingredient, namely 
an UV regularization scheme. Adopting a particularly convenient UV scheme we shall see that the information 
contained in $\EA{k}$ is sufficient in order to determine the related bare action $\BA{\Lambda}$ in the limit 
$\Lambda\rightarrow\infty$.
We prescribe a trajectory $\{\EA{k}, 0\leq k<\infty\}$ and deduce from it how the bare coupling constants contained in 
$\BA{\Lambda}$ must behave in the UV limit when the path integral (with the measure $\mathcal{D}_{\Lambda}\gamma$ and 
action $\BA{\Lambda}$ defined according to the special scheme adopted) is required to reproduce the prescribed $\EA{k}$ 
trajectory. 

Under conditions that we shall spell out precisely later on, one finds that for $\Lambda\rightarrow\infty$
the bare action equals essentially $\Gamma_k$ at $k=\Lambda$:
\begin{equation}\label{uno}
\BA{\Lambda}=\EA{k=\Lambda}+ A_{\Lambda}
\end{equation}
The $\Lambda$-dependent ``correction term" $A_{\Lambda}$ depends on the UV regularization scheme chosen and cannot be found 
from the flow equation. We are going to discuss its general properties and compute it explicitly for various examples, 
including the Einstein-Hilbert truncation of QEG.

Equation (\ref{uno}) is to be regarded a precise, regularized version of the ``rule of thumb" which is quoted frequently,
``$\EA{\infty}=S$". 
In most applications of the EAA in particle and condensed matter physics the $A_{\Lambda}$-contribution in
(\ref{uno}) is completely unimportant and usually not considered explicitly. In fact, in a perturbatively renormalizable theory
 the only effect of the $A_{\Lambda}$-term is to shift those (very few) bare couplings which are relevant at the
Gaussian fixed point. In typical EAA applications one is not interested in their exact values (more precisely, the way how they diverge 
for $\Lambda\rightarrow\infty$) since one anyhow wants to parameterize the RG trajectory by their renormalized counterparts,
to be determined experimentally.

An example in which $A_{\Lambda}$ has been studied in detail  is Liouville field theory \cite{liouv}. There the exact 
values of the bare couplings are of some interest since, being ``almost topological", the RG effects in this theory are so weak
that the running couplings change by only a finite amount during an infinitely long RG time.
In asymptotically safe theories $A_{\Lambda}$ is important for an analogous reason. In fact, the RG trajectories of
Liouville theory cross over from an UV to an IR fixed point so that, in a sense, this theory is asymptotically safe, too \cite{liouv}.

There are various motivations for trying to construct a path integral representation of QEG: 

{\bf{(i)}} The most important motivation, at least from a conceptual point of view, is probably the following.
 In our approach the primary definition of the quantum field theory is in terms of an EAA-trajectory with a 
UV fixed point.
Its endpoint is the ordinary effective action $\Gamma_{k=0}$, so we can easily compute all Green's
functions. However, what we have no easy access to is the microscopic (or ``classical'') system whose standard
quantization gives rise to this particular effective action. A functional integral
representation of the asymptotically safe theory will allow the ``reconstruction" of the microscopic degrees 
of freedom that we implicitly  integrated out in solving the FRGE, as well as their fundamental
interactions. The path integral provides us with their action, and from this action, by a kind of generalized
Legendre transformation, we can reconstruct their Hamiltonian description. From this phase space 
formulation we can read off the classical system whose quantization (also by other methods, canonically say)
leads to the given effective action. We expect this system to be rather complicated so that it cannot be 
guessed easily. This is why we start at the effective level where we know what to look for, namely
a $\Gamma$ whose functional derivatives ($S$-matrix elements) are such that observable quantities have no divergences
on all momentum scales.

{\bf{(ii)}} Another motivation is that many general properties of a quantum field theory are most easily analyzed in a path integral setting,
the implementation of symmetries, the derivation of Ward identities or the incorporation of constraints, to mention just a few. 

{\bf{(iii)}} Many approximation schemes (perturbation theory, large-N expansion, etc.) are more naturally described in a path integral
 rather than a FRGE language. A standard way of doing perturbation theory is to compute, order by order, the counter terms to be included in
  $\BA{\Lambda}$ to get finite physical results in the limit $\Lambda\rightarrow\infty$. Now, QEG
is not renormalizable in perturbation theory and hence new counter terms with free coefficients must be introduced at each order.
If, on the other hand, QEG is asymptotically safe,
defined by a complete trajectory $\{\EA{k}, 0\leq k<\infty\}$, this trajectory ``knows" 
 the correct UV completion of the perturbative calculation. But in order to 
extract this information from $\Gamma_k$ and make contact with the
perturbative language of $S_{\Lambda}$-counter terms  we must convert the $\EA{k}$-trajectory to a $\BA{\Lambda}$-trajectory
first.

{\bf{(iv)}} As a last motivation we mention
that ultimately we would like to understand how QEG relates to other approaches to quantum gravity, such as 
canonical quantization, loop quantum gravity \cite{A,R,T} or Monte Carlo simulations \cite{hamber}-\cite{ajl34}, in which the bare action often
plays a central role. In the Monte Carlo simulations of the Regge and dynamical triangulations formulation, for instance,
the starting point is a regularized path integral involving some discrete version of $\BA{\Lambda}$, and in order to take
the continuum limit one  must fine tune the bare parameters in $\BA{\Lambda}$ in a suitable way.
If one is interested in the asymptotic scaling, for instance, and wants to compare the analytic QEG predictions to the way
the continuum is approached in the simulations, one should convert the $\EA{k}$-trajectory to a $\BA{\Lambda}$-trajectory first.
Note that the map from $\Gamma_k$ to $S_{\Lambda}$, i.e. the associated functional $A_{\Lambda}$ depends explicitly on how precisely the path integral is discretized; 
each alternative formulation of QEG has its own $A_{\Lambda}$!

The remaining sections of this paper are organized as follows. In Section \ref{eaaUV} we discuss the EAA
technology needed later on, the FRGE with a UV cutoff, the relation of its solutions to those of the
$\Lambda$-free FRGE, and we explain why the EAA approach, despite its obvious similarity with the 
Kadanoff-Wilson momentum shell integration is not completely equivalent to it. Then, in Section 
\ref{reconstructing} we demonstrate that every $\Gamma_{k}$-trajectory induces a trajectory of bare actions
and show how it can be found. Section \ref{COSMO} illustrates the method by means of a simple toy model 
which, however, is of physical interest in its own right: the running cosmological constant induced by
a scalar matter field. Section \ref{QEGtruncation} is devoted to QEG. Within the Einstein-Hilbert truncation 
we compute and analyze the map from the effective to the bare couplings in explicit form. In Section
\ref{sectionsix} we give a brief summary and discuss various general conceptual issues related to the
Asymptotic Safety program in QEG.


\section{Effective Average Action with UV cutoff}\label{eaaUV}

In this section we describe how the functional integral underlying the definition of the effective average action  can be made well defined. 
We regularize it by introducing an UV cutoff $\Lambda$ and then derive, in a completely well defined way, the corresponding 
EAA and its flow equation in presence of $\Lambda$. Many different regularization schemes are conceiveable here.
For concreteness we use a kind of ``finite mode regularization"
 which is
ideally suited for implementing the ``background independence" mandatory in QEG.

\subsection{The EAA framework}\label{eaaframework}

In this section, for notational simplicity, we consider a single scalar field on flat space. 
The generalization to more
complicated theories can be achieved  by obvious notational changes.

Let $\chi(x)$ be a real scalar field on a flat $d$-dimensional Euclidean spacetime. In order to discretize momentum space 
we compactify spacetime to a $d$-torus. As a result, the eigenfunctions of the Laplacian 
$\Box=\delta^{\mu\nu}\partial_{\mu}\partial_{\nu}\equiv -\hat{p}^2$ are plane waves $u(x)\propto\exp{(ip\cdot x)}$
with  discrete momenta $p_{\mu}$ and eigenvalues $-p^2$. Given a UV cutoff scale $\Lambda$, there are only finitely
many eigenfunctions with  $|p|\equiv \sqrt{p^2}\leq\Lambda$. We regularize the path integral in the 
UV by restricting the integration to those modes.

As is standard in the EAA construction \cite{avact,avactrev}, the IR modes with $|p|<k$ are suppressed in the path integral by the factor
$\exp(-\Delta_k S[\chi])$ where the functional $\Delta_k S[\chi]$ provides a kind of momentum dependent mass term:
\begin{equation}\label{IR}
\Delta_k S[\chi]=\frac{1}{2}\vol \chi(x)\mathcal{R}_k(\hat{p}^2)\chi(x)
\end{equation}
Now we define a UV-regulated analogue of the standard functional $W_k[J]$:
\begin{equation}\label{wkl}
\exp{\Big(W_{k,\Lambda}[J]\Big)}\equiv \int\mathcal{D}_{\Lambda}\chi\exp\Big( -\BA{\Lambda}[\chi]-\Delta_k S[\chi]
+\vol J(x)\chi(x)\Big)
\end{equation}
The notation in eq.(\ref{wkl}) is symbolic. In fact, its RHS involves only finitely many integrations and is not a genuine
functional integral. The field $\chi$ and the source $J$ in (\ref{wkl}) are ``coarse grained" in the sense that they have an expansion
\begin{equation}
\chi(x)=\sum_{|p|\in[0,\Lambda]}\chi_p\; u_p(x)
\end{equation}
and similar for $J$.
Likewise, the measure $\mathcal{D}_{\Lambda}\chi$ stands for an integration over the Fourier coefficients $\chi_p$ with $p^2$
below $\Lambda^2$:
\begin{equation}\label{measure}
\int\mathcal{D}_{\Lambda}\chi=\prod_{|p|\in[0,\Lambda]}\int_{-\infty}^{\infty}\!\!\textrm{d}\chi_{p}\;M^{-[\chi_{p}]}
\end{equation}
The arbitrary mass parameter $M$ was introduced in order to give the canonical dimension zero to (\ref{measure}).
Even though in eq. (\ref{wkl}) and similar formulas we keep using the familiar (functional) notation, it is to be kept in mind
that $\chi\equiv\{\chi_{p}\}_{|p|\leq\Lambda}$ and $J\equiv\{J_{p}\}_{|p|\leq\Lambda}$ stand for a finite set of variables.

In (\ref{wkl}) the bare action $\BA{\Lambda}$ is allowed to depend on the UV cutoff. Ultimately we would  like to fix
this $\Lambda$-dependence in such a way that, for every finite $k$ and $J$, the path integral has a well defined limit for
$\Lambda\rightarrow\infty$.

Denoting the Legendre transform of $W_{k,\Lambda}[J]$ with respect to $J$ by $\widetilde{\Gamma}_{k,\Lambda}[\phi]$ the EAA
is defined as \cite{avact}
\begin{equation}\label{effectiveaction}
\EA{k,\Lambda}[\phi]\equiv \widetilde{\Gamma}_{k,\Lambda}[\phi]-\frac{1}{2}\vol \phi(x)\mathcal{R}_{k}(\hat{p}^{2})\phi(x)
\end{equation}
Here $\phi=\{\phi_{p}\}_{|p|\in[0,\Lambda]}$ is the expectation value field $\phi(x)\equiv \langle \chi(x)\rangle$ obtained by 
differentiating $W_{k,\Lambda}$. In the usual notation,
\begin{equation}\label{phi}
\phi(x)=\frac{\delta}{\delta J(x)}W_{k,\Lambda}[J]
\end{equation}
If this relation can be inverted in the form $J(x)=\mathcal{J}_{k,\Lambda}[\phi](x)$ we have
\begin{equation}\label{effectiveaction1}
\widetilde{\Gamma}_{k,\Lambda}[\phi]=\vol \;\phi(x)\;\mathcal{J}_{k,\Lambda}[\phi](x)-W_{k,\Lambda}[\mathcal{J}_{k,\Lambda}[\phi]]
\end{equation}
and
\begin{equation}\label{J}
\frac{\delta}{\delta\phi(x)}\widetilde{\Gamma}_{k,\Lambda}[\phi]=\mathcal{J}_{k,\Lambda}[\phi](x)
\end{equation}

By following the usual steps \cite{avact} it is straightforward to show that the definition (\ref{effectiveaction}) implies the following
exact FRGE for $\EA{k,\Lambda}$:
\begin{equation}\label{frge}
k\partial_{k}\EA{k,\Lambda}[\phi]=\frac{1}{2}\textrm{Tr}_{\Lambda}\Big[ \Big(\EA{k,\Lambda}^{(2)}[\phi]+\mathcal{R}_{k}\Big)^{-1}k\partial_{k}\mathcal{R}_{k}\Big]
\end{equation}
Here, $\textrm{Tr}_{\Lambda}$ denotes the trace restricted to the subspace spanned by the eigenfunctions of $p^{2}$ with eigenvalues smaller than $\Lambda^{2}$:
\begin{equation}\label{trace}
\textrm{Tr}_{\Lambda}[\cdots]=\textrm{Tr}\Big[\theta(\Lambda^{2}-\hat{p}^{2})[\cdots]\Big]
\end{equation}
As it is customary, $\EA{k,\Lambda}^{(2)}$ denotes the Hessian of $\EA{k,\Lambda}$, interpreted as an operator constructed from $\hat{p}_{\mu}$ and the conjugate position variable
$\hat{x}^{\mu}$. 

Note that (\ref{frge}) can be rewritten in the form
\begin{equation}\label{frge1}
k\frac{\partial}{\partial k}\EA{k,\Lambda}[\phi]=\frac{1}{2}\;k\frac{D}{D k}\;\textrm{Tr}_{\Lambda}\ln{\Big[\EA{k,\Lambda}^{(2)}[\phi]+
\mathcal{R}_{k}\Big]}
\end{equation}
where the derivative $D/D k$ acts on the $k$ dependence of $\mathcal{R}_{k}$ only. From (\ref{frge1}) we can read off the the 1-loop approximation to the solution of the
FRGE:
\begin{equation}\label{frge2}
\EA{k,\Lambda}[\phi]\approx \frac{1}{2}\textrm{Tr}_{\Lambda}\Big[\BA{k,\Lambda}^{(2)}[\phi]+
\mathcal{R}_{k}\Big] + \mathcal{S}_{\Lambda}[\phi]
\end{equation}
The constant of integration $\mathcal{S}_{\Lambda}$ is  related to, but not equal to the bare action $\BA{\Lambda}$. 

It can be shown using 
(\ref{effectiveaction}), (\ref{phi}), (\ref{effectiveaction1}) and (\ref{J}) that $\EA{k,\Lambda}$ satisfies the following 
integro-differential equation:
\begin{eqnarray}\label{integroeq0}
\exp{\Big(-\EA{k,\Lambda}[\phi]\Big)}& = & \int\!\!\mathcal{D}_{\Lambda}\chi \exp{\Big(-\BA{\Lambda}[\chi]+    \vol(\chi-\phi)\frac{\delta\EA{k,\Lambda}[\phi]}{\delta \phi}}
\nonumber\\
& & {} -\frac{1}{2}\vol (\chi-\phi) \mathcal{R}_{k}(\hat{p}^{2})(\chi-\phi)\Big)
\end{eqnarray}
In terms of the fluctuation field $f(x)\equiv \chi(x)-\phi(x)$ it reads,
\begin{eqnarray}\label{integroeq}
\exp{\Big(-\EA{k,\Lambda}[\phi]\Big)}& = &\int\!\!\mathcal{D}_{\Lambda}f \exp{\Big(-\BA{\Lambda}[\phi+f]+    \vol f(x) \frac{\delta\EA{k,\Lambda}[\phi]}{\delta \phi(x)}}
\nonumber\\
& & {}-\frac{1}{2}\vol f(x)\mathcal{R}_{k}(\hat{p}^{2})f(x)\Big)
\end{eqnarray}

\subsection{EAA vs. momentum shell integration}\label{eaamomentum}

\begin{figure}
\centering
\begin{tabular}{ll}
\bf{(a)} & \bf{(b)}\\
\includegraphics[width=7.3cm,height=5cm]{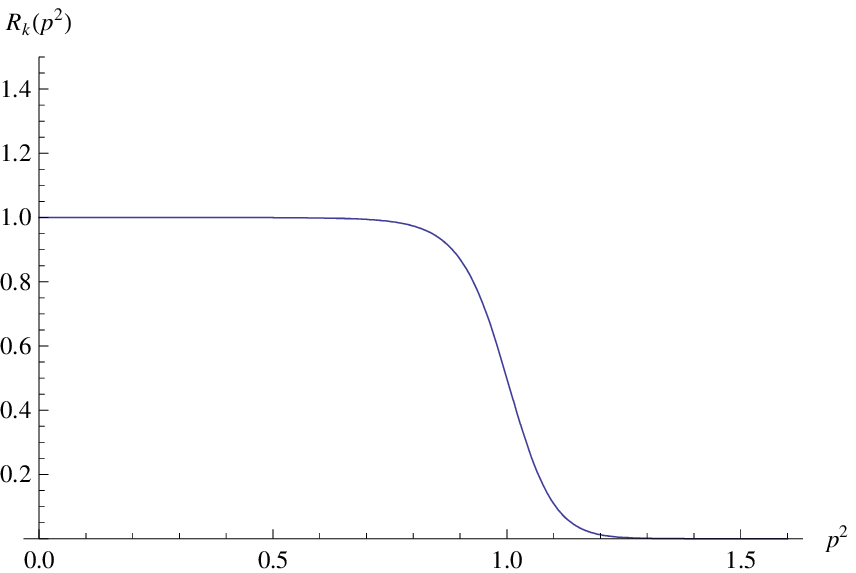} & \includegraphics[width=7.3cm,height=5cm]{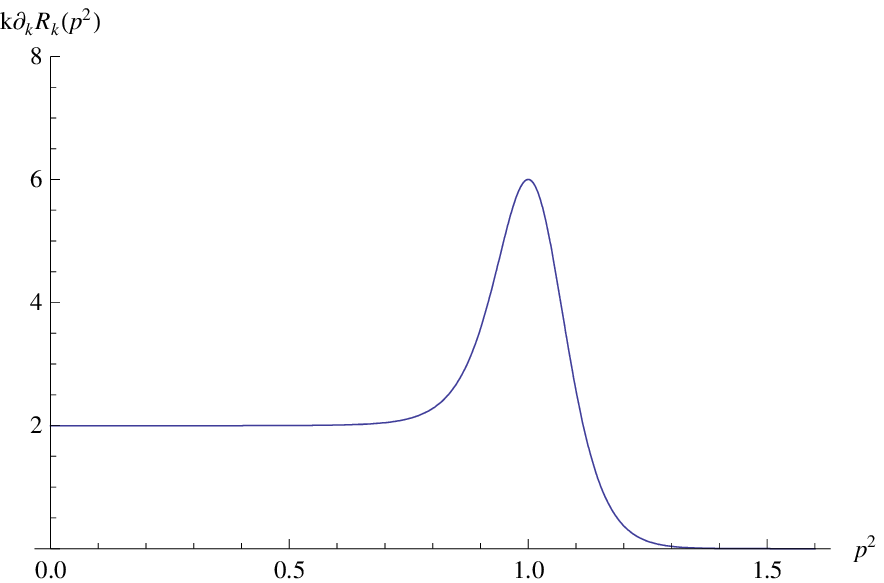}
\end{tabular}
\caption{\small{A typical non-singular cutoff function {\bf{(a)}} and its scale
derivative {\bf{(b)}}. Dimensionful quantities are expressed in units of $k$.}}
\end{figure}

The key feature of the EAA is the mode suppression term $\Delta_{k}S$ which gives a mass of order $k$ to the field modes with momenta $p\leq k$.
How this happens precisely is controlled by the function $\mathcal{R}_{k}({p}^{2})$. The details of this function are irrelevant to a large extent;
we only require that $\mathcal{R}_{k}({p}^{2})$ is a monotonic function of $p^{2}$ which interpolates between $\mathcal{R}_{k}({p}^{2}\rightarrow 0)=k^2$ and 
$\mathcal{R}_{k}({p}^{2}\rightarrow\infty)=0$, whereby the transition between the two regimes takes place near $p^{2}=k^{2}$. 
An example sketched in Fig.1a. Its scale derivative $k\partial_{k}\mathcal{R}_{k}({p}^{2})$  has a peak near $p^{2}=k^{2}$, a very rapid  (exponential) decay for $p^{2}\gg k^{2}$, and for $p^{2}\ll k^{2}$ a plateau on which $k\partial_{k}\mathcal{R}_{k}({p}^{2})$
is an approximately constant function of $p^{2}$. We shall refer to the $p^{2}\gg k^{2}$ and  $p^{2}\ll k^{2}$ regime of $k\partial_{k}\mathcal{R}_{k}({p}^{2})$,
respectively, as the ``exponential tail'' and the ``low-$p$ continuum''.

Most of the  somewhat unexpected features of the EAA that we are going to discuss in this paper are due to the
``low-$p$ continuum''. It owes its existence to the specific way the EAA, with a non-singular $\mathcal{R}_{k}$, treats the modes with $p^{2}<k^{2}$.
Rather than excising  them completely from the functional integral (as done for the UV modes with $p^{2}>\Lambda^{2}$) they are only weakly suppressed by means of a mass term
$\mathcal{R}_{k}\approx k^{2}$; it is essentially constant for $p^{2}\ll k^{2}$ and hence yields the plateau value   $k\partial_{k}\mathcal{R}_{k}\approx 2k^{2}$
for its scale derivative, see Fig.1.

The advantage of this very smooth IR suppression, and in fact its main motivation, are the regularity properties it entails for the resulting EAA.
Its disadvantage is that it complicates the interpretation  to some extent since the EAA with this type of cutoff is {\it{not}} in accord with the familiar 
picture of a ``momentum shell integration'' which is often used in the standard formulations of the Wilsonian renormalization group. If one wants to literally mimic 
a momentum shell integration within the EAA framework one would have to give to $\mathcal{R}_{k}(p^{2})$ a singular profile such that $k\partial_{k}\mathcal{R}_{k}$ 
is sharply peaked near $p^{2}=k^{2}$ and vanishes rapidly for both $p^{2}\ll k^{2}$ and $p^{2}\gg k^{2}$. In this case, the trace in (\ref{frge}) would receive 
contributions from a thin shell of eigenvalues near $p^{2}=k^{2}$ only, consistent with the standard Wilson-Kadanoff picture.

If  one uses a non-singular $\mathcal{R}_{k}$ like the one in Fig.1 the trace on the RHS of the FRGE can receive contributions
from {\it{all}} modes with momenta below $k$. Whether this has a qualitatively important impact on the RG running of the generalized couplings parameterizing $\Gamma_{k,\Lambda}$ depends on which couplings
are considered and, in practice, on the truncation. To explain this point we assume that the EAA is expanded in terms of field monomials $I_{\alpha}[\phi]$ as
$\EA{k,\Lambda}[\phi]=\sum_{\alpha}g_{\alpha}(k,\Lambda)I_{\alpha}[\phi]$ or as a Volterra series involving its $n$-point functions $\EA{k,\Lambda}^{(n)}(x_{1},\cdots,x_{n})$. Then
 we can find the $\beta$-functions of the coupling $g_{\alpha}$ or the $n$-point functions by repeatedly differentiating the FRGE (\ref{frge}) and 
setting $\phi=0$ thereafter. If one computes $\textrm{Tr}_{\Lambda}$ in momentum space, this leads to a representation of
most $\beta$-functions in terms of integrals which contain products of the modified propagators
\begin{equation}\label{propagator}
\frac{1}{\EA{k,\Lambda}^{(2)}[0](p)+\mathcal{R}_{k}(p^{2})}
\end{equation}
as well as the vertices implied by $\EA{k,\Lambda}$. The Feynman diagrams summed up in this way are 
similar to those of standard perturbation theory. For the $\beta$-functions which indeed do have this structure the term $\mathcal{R}_{k}(p^{2})$ in (\ref{propagator})
acts as an IR regulator: It equips the low-$p$ modes with a non zero mass $\mathcal{R}_{k}(p^{2})\approx k^{2}$, thus suppressing their contribution inside loops.

While this argument applies to most couplings, there are also exceptions. They arise in the computation of those $\beta$-functions which can be projected
out of the RHS of the flow equation by acting with only very few derivatives $\delta/\delta\phi$ on it, or  with no derivatives at all. In the exceptional cases
the impact of the mass $\mathcal{R}_{k}(p^{2})$ is paradoxical in the sense that it does not lead to a suppression of 
the small-$p$
modes but rather to their {\it{enhancement}}.

To illustrate how this can happen let use the 1-loop formula (\ref{frge2}) in order to determine the $k$-dependence of the cosmological constant induced by the
scalar. It obtains by setting $\phi=0$ directly in (\ref{frge2}), without performing any derivative:
\begin{equation}\label{loopzero}
\frac{1}{2}\int_{|p|<\Lambda}\!\!\frac{\textrm{d}^{d}p}{(2\pi)^{d}}\ln{\big(p^{2} +\mathcal{R}_{k}(p^{2})\big)}
\end{equation}
For simplicity we assumed a free massless theory here, with $\EA{k,\Lambda}^{(2)}[0]=p^{2}$. The integrand of (\ref{propagator}) equals $\ln{(p^{2}+k^{2})}$ for
$|p|\lesssim k$ and $\ln{(p^{2})}$ for $k\lesssim |p|<\Lambda$. As a result, 
the $k$-dependence of (\ref{loopzero}) is entirely due to the former regime, and
the integral is an {\it{increasing}} function of $k$ therefore.
Thus, the higher is the IR cutoff $k$ the larger is the contribution of the low-$p$ modes to the cosmological constant. This is the paradoxical effect we mentioned:
instead of suppressing the contribution of the IR modes to the running of the couplings, the addition of $\Delta_{k}S$ leads to an enhancement here.

Under appropriate conditions (perturbation theory, perturbatively renormalizable model, etc.) the Appelquist-Carrazzone decoupling theorem \cite{applecarra}
tells us in which way a particle whose mass is made very heavy ``disappears'' from the theory: if the remaining theory without this particle is renormalizable, the heavy particle 
manifests itself either via a {\it{renormalization of its relevant coupling constants}} or by effects that are suppressed by inverse powers of its mass.

While strictly speaking the theorem of perturbation theory cannot be applied literally in the broader context envisaged here we would nevertheless expect the ``paradoxical''  
enhancement in the large $k$-limit to occur for a small set of ``relevant'' parameters, while the $\beta$-functions of the ``irrelevant'' ones show the ordinary
decoupling behavior, i.e. they vanish at large $k$.

\subsection{Removing the UV cutoff from the FRGE}\label{removingUV}

In the following we assume that the cutoff is chosen such that $k\partial_k\mathcal{R}_{k}(p^{2})$ decreases as a function of 
$p^2$, at $p^2\gg k^2$, sufficiently rapidly for the trace on the RHS of the flow equation to exist even in the 
limit when the UV cutoff is removed, $\Lambda\rightarrow \infty$.
The resulting ``$\Lambda$-free" FRGE without UV cutoff, valid for all $k\geq 0$, has the familiar form:
\begin{equation}\label{fluxeq1}
k\partial_k\EA{k}[\phi]=\frac{1}{2}\textrm{Tr}\Big[ \Big(\EA{k}^{(2)}[\phi] + 
\mathcal{R}_{k}(p^{2})\Big)^{-1}k\partial_k\mathcal{R}_{k}(p^{2})\Big]
\end{equation}
A {\it{complete}} solution of (\ref{fluxeq1}) is a family
of functionals $\EA{k}[\phi]$ defined for any value of $k\in [0,\infty)$.
Later on it will be convenient to write the $\Lambda$-free FRGE as
\begin{equation}
k\partial_k\EA{k}[\phi]=B_k\{\EA{k}\}[\phi]
\end{equation}
where $B_k$ denotes the ``beta functional"
\begin{equation}\label{bk}
B_k\{\Gamma\}[\phi]=\frac{1}{2}\textrm{Tr}\Big[ \Big(\Gamma^{(2)}[\phi] + 
\mathcal{R}_{k}(p^{2})\Big)^{-1}k\partial_k\mathcal{R}_{k}(p^{2})\Big]
\end{equation}
Actually the map $B_k$ is a kind of ``hyperfunctional" of its argument $\Gamma$ and an ordinary functional of $\phi$.
Geometrically speaking it describes a vector field on theory space.

\subsection{$\EA{k}$ vs. $\EA{k,\Lambda}$ in the limit $\Lambda\rightarrow\infty$}\label{section4}

It is natural to ask how solutions $\EA{k}$ of the $\Lambda$-free flow equation (\ref{fluxeq1}) relate to solutions
$\EA{k,\Lambda}$ of the original FRGE (\ref{frge}) in the limit where $\Lambda$ becomes large. To answer this question we 
compare the vector fields driving the RG evolution of $\EA{k}$ and $\EA{k,\Lambda}$, respectively.

The FRGE with UV cutoff, eq (\ref{frge}), contains the restricted trace $\textrm{Tr}_{\Lambda}$ of (\ref{trace}). 
Rewriting the latter as
\begin{equation}\label{trace2}
\textrm{Tr}_{\Lambda}[\cdots]=\textrm{Tr}[\cdots]-\textrm{Tr}[\theta(\hat{p}^2-\Lambda^2)(\cdots)]
\end{equation}
implies the following representation of the RG equation:
\begin{equation}\label{fluxeq2}
k\partial_k\EA{k,\Lambda}[\phi]=B_k\{\EA{k,\Lambda}\}[\phi]+\Delta B_{k,\Lambda}\{\EA{k,\Lambda}\}[\phi]
\end{equation}
Here $B_k$ is defined as in (\ref{bk}) and the second term on the RHS involves the functional
\begin{equation}\label{deltabk}
\Delta B_{k,\Lambda}\{\EA{}\}[\phi]\equiv -\frac{1}{2}\textrm{Tr}\Big[ \theta(\hat{p}^2-\Lambda^2)\;\Big(\EA{}^{(2)} + 
\mathcal{R}_{k}\Big)^{-1}\;k\partial_k\mathcal{R}_{k}\Big]
\end{equation} 
The first term on the RHS of  (\ref{fluxeq2}) is the same as in the $\Lambda$-free FRGE, the second is a correction to the
beta functional due to the UV cutoff; it affects $\EA{k, \Lambda}$ but not $\EA{k}$. The corresponding RG flows are
generated by the vector fields $B_k +\Delta B_{k,\Lambda}$ and $B_k$, respectively.

The term $\Delta B_{k,\Lambda}$ is ``small" in the following sense. Thanks to the step function under the trace of (\ref{deltabk})
the latter receives contributions only from modes with eigenvalues $p^2>\Lambda^2\geq k^2$. However, for $p^2$ larger than $k^2$
the last factor under the trace, $k\partial_k\mathcal{R}_k$, decays very quickly when $p^2\rightarrow \infty$. As a result,
only very few modes can give a substantial contribution to $\Delta B_{k,\Lambda}$, and this contribution diminishes quickly when 
$\Lambda\rightarrow\infty$ at fixed $k$. 

This argument shows that the flow equations for $\EA{k}$ and $\EA{k,\Lambda}$ are
essentially the same as long as $k\ll \Lambda$. When $k$ approaches $\Lambda$ from below, small deviations will occur due to
$\Delta B_{k,\Lambda}$. Making $\Lambda$ larger the range of $k$-values in which $\EA{k}$ and $\EA{k, \Lambda}$ have the same beta 
functional expands, and finally, in the limit $\Lambda\rightarrow\infty$, $\EA{k}$ and $\EA{k,\Lambda}$ have the same
scale derivatives at any finite $k$.

This is the situation for a generic non-singular $\mathcal{R}_k$. It is very convenient that there exists actually a special cutoff, 
the optimized cutoff \cite{opt}, for which the correction term $\Delta B_{k,\Lambda}$ vanishes identically:
\begin{equation}\label{bk0}
\Delta B_{k,\Lambda}^{\rm{opt}}=0 \quad \forall\;\; k\leq\Lambda
\end{equation} 
The optimized cutoff is given by
\begin{equation}\label{optcutoff}
\mathcal{R}_k(p^2)=(k^2-p^2)\theta(k^2-p^2)
\end{equation}
which entails
\begin{equation}\label{optcutoff1}
k\partial_k\mathcal{R}_k(p^2)=2k^2\theta(k^2-p^2)
\end{equation}
With (\ref{optcutoff1}), the trace defining $\Delta B_{k,\Lambda}$ contains a factor of $\theta(\hat{p}^2-\Lambda^2)
\theta(k^2-\hat{p}^2)$
and therefore it vanishes identically since $k\leq\Lambda$.

When the optimized cutoff is used the relationship between the solutions of the flow equations with and without an UV 
cutoff are easy to describe:

For all $k\leq\Lambda$ the functional $\EA{k,\Lambda}$ satisfies the same FRGE as $\EA{k}$, namely the $\Lambda$-free flow equation $k\partial_k\EA{k,\Lambda}=B_k\{\EA{k,\Lambda}\}$. Therefore, if we impose at some arbitrary scale $k_1\leq\Lambda$
the same initial conditions on both functionals,
$
\EA{k_1,\Lambda}=\EA{\textrm{initial}}=\EA{k_1}
$
the solutions of the two flow equations agree exactly in the range of $k$-values in which both of them are defined,
i.e. for $k\leq\Lambda$:
\begin{equation}\label{inicond0}
\EA{k,\Lambda}=\EA{k}\quad\textrm{ when }0\leq k \leq\Lambda
\end{equation}
In particular, this relationship holds true at $k=\Lambda$:
\begin{equation}\label{inicond1}
\EA{\Lambda,\Lambda}=\EA{\Lambda}
\end{equation}
In  (\ref{inicond0}) and (\ref{inicond1}), $\Lambda$ is a fixed, but arbitrary finite scale.

Let us assume we have solved the  $\Lambda$-free FRGE and found some {\it{complete}} solution
\begin{equation}\label{solution}
\{\EA{k},\; 0\leq k <\infty\}
\end{equation}
Being complete means that it extends from $k=0$ to $k``="\infty$, i.e. it has a well defined IR and UV limit, respectively.
In the case we are mostly interested in, QEG, the (dimensionless form of) $\EA{k}$ runs into a fixed point for 
$k\rightarrow\infty$ so that it has indeed a well defined UV limit. Knowing the solution (\ref{solution}) we immediately know by 
(\ref{inicond0}) also a  solution to the FRGE {\it{with an UV cutoff}}, namely 
\begin{equation}\label{solutionLambda}
\{\EA{k,\Lambda},\; 0\leq k <\Lambda\}
\end{equation}  
where $\EA{k,\Lambda}$ equals the $\EA{k}$ of (\ref{solution}) for $k$ below $\Lambda$.

Since $\Lambda$ is arbitrary we can make it as large as we like, in particular we can take the limit
\begin{equation}
\lim_{\Lambda\rightarrow\infty}\EA{k,\Lambda}\equiv \EA{k,\infty}\quad\textrm{ with }k<\infty\textrm{ fixed.}
\end{equation}
Thanks to eq.(\ref{inicond0}) this limit does indeed exists and is given by
\begin{equation}
\EA{k,\infty}=\EA{k}\quad\textrm{ for all }k\geq0
\end{equation}
\begin{figure}\label{figure2}
\centering
\includegraphics[width=10cm,height=7cm]{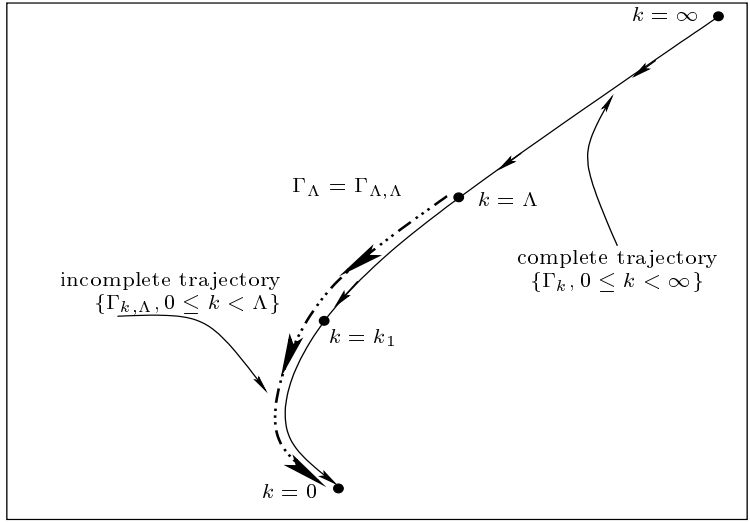} 
\caption{\small Employing the optimized cutoff every complete solution to the $\Lambda$-free FRGE gives rise to a  solution
of the FRGE with UV cutoff, valid up to any  value of $\Lambda$. }
\end{figure}

The situation with the optimized cutoff is sketched in Fig.2. Here, {\it{$\EA{k,\Lambda}$ is simply the restriction of\  $\EA{k}$
\ to the interval $k\leq\Lambda$}}. For a generic cutoff, the trajectories $\EA{k,\Lambda}$ and $\EA{k}$ passing
through the same $\EA{\textrm{initial}}$ differ slightly when $k$ approaches $\Lambda$ but the qualitative picture is 
similar.

\subsection{An illustrative example: the local potential \\ 
approximation}\label{LPA}

In order to illustrate the above reasoning we shall now consider approximate solutions to the flow equation 
(\ref{frge}) on the truncated theory space spanned by actions of the form
\begin{equation}\label{actionLPA}
\EA{k,\Lambda}[\phi]=\vol\Big\{ \frac{1}{2}\partial_{\mu}\phi\partial^{\mu}\phi + U_{k,\Lambda}(\phi)\Big\}
\end{equation} 
This truncation is referred to
as the local potential approximation (LPA). Inserting the ansatz (\ref{actionLPA}) into the FRGE (\ref{frge})
we obtain a partial differential equation for the potential $U_{k,\Lambda}$. The projection on the non-derivative part
of the action can be performed  by inserting a constant field $\phi(x)=\textrm{const}\equiv \phi$.
Then the trace is easily evaluated in a plane wave basis, with the result (for $k\leq \Lambda$)
\begin{equation}\label{RGpotential}
k\partial_k U_{k,\Lambda}(\phi)=v_d\int_0^{\Lambda^2}\!\!\textrm{d}y\; y^{(d-2)/2}\;
\frac{k\partial_k\mathcal{R}_k(y)}{y+\mathcal{R}_k(y)+U^{\phantom{1}''}_{k,\Lambda}(\phi)}
\end{equation}
Here $v_d\equiv [2^{d+1}\pi^{d/2}\Gamma(d/2)]^{-1}$, and $y\equiv p^2$ is the square of radial coordinate
in momentum space. The RG equation (\ref{RGpotential}) nicely illustrates the two different ways
in which the UV and IR cutoffs, respectively, are implemented:

The {\bf{UV cutoff}} built into the {\it{measure}} has led to a {\it{sharp}} restriction of the interval the momenta
 are integrated over: $y\equiv p^2\in[0,\Lambda^2]$.
The {\bf{IR cutoff}}, instead, consists of a momentum dependent mass term introduced into the {\it{action}}; rather 
than delimiting the $p^2$-integration it affects the mode sum (the integral in (\ref{RGpotential})) in a
{\it{smooth}} way only, via the function $\mathcal{R}_k$ in the integrand.

The integral representation (\ref{RGpotential}) is valid for any choice of $\mathcal{R}_k$. For a generic one, the integral does indeed have a (weak, see above) dependence on $\Lambda$. As expected, the situation is particularly simple for the optimized cutoff (\ref{optcutoff}). In this case we can perform the $y$-integral in closed form and obtain for all $k\leq\Lambda$:
\begin{equation}\label{RGpotential1}
k\partial_k U_{k,\Lambda}(\phi)=\frac{4v_d}{d}\;\frac{k^{d+2}}{k^2+
U^{\phantom{1}''}_{k,\Lambda}(\phi)}
\end{equation}
We see that this equation has no {\it{explicit}} dependence on $\Lambda$ at all.

With the $\Lambda$-free FRGE (\ref{fluxeq1}) we can proceed analogously. Upon inserting a LPA similar to (\ref{actionLPA}),
this time for $\EA{k}$ and with a potential $U_k(\phi)$, we obtain a RG equation which coincides with (\ref{RGpotential})
except that the upper limit of integration, $\Lambda^2$, is now replaced by infinity. For the special case of the optimized
cutoff it implies
\begin{equation}\label{RGpotential2}
k\partial_k U_{k}(\phi)=\frac{4v_d}{d}\;\frac{k^{d+2}}{k^2+U^{''}_{k}(\phi)}
\end{equation}
Eq.(\ref{RGpotential2}) has exactly the same structure as (\ref{RGpotential1}). However, the equation for 
$\EA{k}$, eq.(\ref{RGpotential2}) is valid for all $k>0$, while (\ref{RGpotential1}) for $\EA{k,\Lambda}$ holds 
true in the interval $k\in[0,\Lambda]$ only. Thus we see explicitly that if we impose the same initial conditions in both 
cases, the respective solutions are related by
\begin{equation}
U_{k,\Lambda}(\phi)=U_{k}(\phi)\quad\forall\;\;k\leq\Lambda
\end{equation}
i.e. $U_{k,\Lambda}(\phi)$ is  the restriction of $U_{k}(\phi)$ to the $k$-values smaller than $\Lambda$.


\section{Reconstructing the bare action}\label{reconstructing}

Our key requirement is that the functionals $W_{k,\Lambda}$ and $\EA{k,\Lambda}$, $k\leq\Lambda$, remain finite in the limit
$\Lambda\rightarrow\infty$. To achieve this, the bare action $\BA{\Lambda}$ must be given a specific $\Lambda$-dependence.
This $\Lambda$-dependence itself depends on the  UV regularization that was chosen  to make the path integral 
well defined. In the explicit examples below we keep using the ``finite mode" regularization for this purpose.

\subsection{The input: $\EA{\Lambda, \Lambda}$}\label{theimput}

The problem we are going to address next is how one can determine the corresponding $\Lambda$-dependence of $\BA{\Lambda}$ if one knows some solution of the $\Lambda$-free flow equation.

Let us assume we are given an exact solution $\{\EA{k},\; k\in [0,\infty)\}$ of the $\Lambda$-free FRGE,
i.e. a complete RG trajectory extending from $k=0$ to $k``="\infty$. By the construction discussed in subsection \ref{section4}
it implies a solution to the FRGE with an UV cutoff:
$\{\EA{k,\Lambda},\; k\in [0,\Lambda]\}$, $\Lambda$ arbitrary, fixed. The trajectories are related by eq.(\ref{inicond0}), and setting 
$k=\Lambda$ we have in particular $\EA{\Lambda,\Lambda}=\EA{\Lambda}$ or, more explicitly, 
\begin{equation}
\EA{k=\Lambda,\Lambda}=\EA{k=\Lambda}
\end{equation}
Thus, knowing $\EA{k}$ for all $k$ means that we know $\EA{\Lambda,\Lambda}$ for all $\Lambda$. Next we shall demonstrate
how, given $\EA{\Lambda,\Lambda}$, the bare action $\BA{\Lambda}$ can be (re)constructed.

\subsection{The saddle point expansion}\label{sectionsaddle}

The desired relation between the bare action and the average action can be deduced from the integro-differential equation
(\ref{integroeq}):
\begin{equation}\label{eqsaddlepoint}
\exp{\Big(-\EA{k,\Lambda}[\phi]\Big)}=\int\!\!\mathcal{D}_{\Lambda}f \exp{\Big(  -\BA{\textrm{tot}}[f;\phi]\Big)}
\end{equation}
Here we set
\begin{equation}
\BA{\textrm{tot}}[f;\phi]\equiv \BA{\Lambda}[\phi+f] -  \vol f(x) \frac{\delta\EA{k,\Lambda}[\phi]}{\delta \phi(x)}-\frac{1}{2}\vol f(x)\mathcal{R}_{k}(\hat{p}^{2})f(x)
\end{equation}
We must ``solve'' eq.(\ref{eqsaddlepoint}) for $\BA{\Lambda}$ in the limit $k=\Lambda\rightarrow\infty$. The obvious problem we encounter here is
that we have to explicitly perform the integration over $f$. The fact which to some extent comes to our rescue here is that we need to know (\ref{eqsaddlepoint})
only for $k=\Lambda$ which implies that {\it{all}} modes contributing to the $f$-integral have a mass of order $\Lambda$ 
and this mass becomes arbitrarily large in the limit
of interest. As a consequence, many of the contributions that could potentially occur in the equation relating $\EA{\Lambda,\Lambda}$ to $\BA{\Lambda}$ will
vanish for $\Lambda\rightarrow\infty$ since the underlying loop integrals involve infinite propagator masses. However, we must worry about those
field monomials or $n$-point functions contained in $\EA{\Lambda,\Lambda}$ on which the IR cutoff $\mathcal{R}_{k}$ has the ``paradoxical'' effect of
enhancing rather than suppressing them. They will diverge for $\Lambda\rightarrow\infty$ and these divergences must be absorbed by $\BA{\Lambda}$.
(In perturbation theory this concerns precisely the terms which are relevant and marginal at the Gaussian fixed point.)

In order to make these ideas explicit we evaluate the $f$-integral by means of a saddle point approximation. We expand $f(x)\equiv f_{0}(x)+h(x)$ where 
$f_{0}$ is the stationary point of $S_{\textrm{tot}}$, i.e. $(\delta S_{\textrm{tot}}/\delta f)[f_{0}]=0$, or explicitly,
\begin{equation}\label{saddlepointcond}
\frac{\BA{\Lambda}}{\delta\phi(x)}[\phi+f_{0}]-\frac{\delta \EA{k,\Lambda}}{\delta\phi(x)}[\phi]+\mathcal{R}_{k}f_{0}(x)=0
\end{equation}
Note that besides being a function of $k$ and $\Lambda$, the saddle point $f_0(x)\equiv f_{0}[\phi](x;k,\Lambda)$ is a functional of $\phi$.
(It will be instructive to allow for arbitrary $k\leq\Lambda$ and set $k=\Lambda$ later only. )

 Here we shall analyze only the leading order of the saddle point 
expansion, i.e. the 1-loop approximation. Upon inserting $f=f_{0}+h$ into (\ref{eqsaddlepoint}), expanding $S_{\textrm{tot}}$ to second order in $h$, and 
performing the Gaussian integral over $h$ we obtain the following relationship between the bare and the average action:
\begin{eqnarray}\label{loopexpansion}
\EA{k,\Lambda}[\phi] &=& S_{\Lambda}[\phi+f_{0}]-\vol f_{0}\frac{\delta\EA{k,\Lambda}[\phi]}{\delta\phi}+\frac{1}{2}\vol f_{0}\mathcal{R}_{k}f_{0}+{}
\nonumber\\
& & {}+  \frac{1}{2}\textrm{Tr}_{\Lambda}\ln{\Big[\Big(\frac{\delta^{2}S_{\Lambda}[\phi+f_{0}]}{\delta\phi^{2}}+\mathcal{R}_{k}\Big)M^{-2}\Big]}+\cdots
\end{eqnarray}
Reinstating $\hbar$ as a loop counting parameter for a moment the $\textrm{Tr}_{\Lambda}\ln{[\cdots]}$ term in (\ref{loopexpansion}) is of order
$\hbar$, while the dots in (\ref{loopexpansion}) stand for $\mathcal{O}(\hbar^2)$ terms which we neglect. The stationary point $f_{0}$, too, has an expansion in powers 
of $\hbar$. To find it we expand the saddle point condition (\ref{saddlepointcond}) for small $f_{0}$:
\begin{equation}\label{rel1}
\Big\{\frac{\delta^{2}S_{\Lambda}[\phi]}{\delta\phi^{2}}+\mathcal{R}_{k}\Big\}f_{0}=\frac{\delta}{\delta\phi}\Big(\EA{k,\Lambda}-S_{\Lambda}\Big)[\phi] +\mathcal{O}(f_{0}^{2})
\end{equation}
Likewise the expansion of (\ref{loopexpansion}) yields, in a symbolic notation,
\begin{eqnarray}\label{rel2}
\EA{k,\Lambda}[\phi]-S_{\Lambda}[\phi] &=& -\int f_{0}\frac{\delta}{\delta\phi}\Big(\EA{k,\Lambda}-S_{\Lambda}\Big)[\phi] + 
\frac{1}{2}\int f_{0}\Big( S^{(2)}_{\Lambda}[\phi]+\mathcal{R}_{k}\Big)f_{0} + \mathcal{O}(f_{0}^{3})+{}
\nonumber\\
 & & {} \!\!\!\!\!\! \!\!\!\!\!\!\!\!\! \!\!\! \!\!\!\!\!\! \!\!\!\!\!\!\!\!\! \!\!\! + \frac{\hbar}{2}\;\textrm{Tr}_{\Lambda}\ln{\Big\{  \big[  S^{(2)}_{\Lambda}[\phi]+S^{(3)}_{\Lambda}[\phi]f_{0}+S^{(4)}_{\Lambda}[\phi]f_{0}f_{0}+
 \cdots+\mathcal{R}_{k}\big]M^{-2}\Big\}}+ \mathcal{O}(\hbar^{2})
\end{eqnarray}

The coupled relations (\ref{rel1}) and (\ref{rel2}) are solved self-consistently if $f_{0}=0+\mathcal{O}(\hbar)$ and $\EA{k,\Lambda}[\phi]-S_{\Lambda}[\phi]=\mathcal{O}(\hbar)$
which leads to the following 1-loop formula for the difference between the average and the bare action:
\begin{equation}
\EA{k,\Lambda}[\phi]-S_{\Lambda}[\phi]=\frac{1}{2}\textrm{Tr}_{\Lambda}\ln{\Big\{ \big[ S^{(2)}_{\Lambda}[\phi]+\mathcal{R}_{k} \big]M^{-2} \Big\} }
\end{equation}
Setting $k=\Lambda$ we arrive at the final result
\begin{equation}\label{finalequation}
\EA{\Lambda,\Lambda}[\phi]-S_{\Lambda}[\phi]=\frac{1}{2}\textrm{Tr}_{\Lambda}\ln{\Big\{ \big[ S^{2}_{\Lambda}[\phi]+\mathcal{R}_{\Lambda} \big]M^{-2} \Big\} }
\end{equation}
Here and in the following we set $\hbar=1$ again.

Equation (\ref{finalequation}) is the desired relation which tells us how $S_{\Lambda}$ must depend on $\Lambda$ in order to give rise
to the prescribed $\EA{\Lambda,\Lambda}$. For every given, fixed $\EA{\Lambda}[\phi]$ eq.(\ref{finalequation}) is a complicated functional differential equation for
$S_{\Lambda}[\phi]$, involving second derivatives $S^{(2)}_{\Lambda}\equiv \delta^{2}S_{\Lambda}/\delta\phi\delta\phi$ under the restricted trace.

The relation (\ref{finalequation}), and its obvious generalizations to more complicated theories, is our main tool for (re)constructing the path integral
that belongs to a known solution of the FRGE.

\subsection{The LPA example}\label{LPAexample}

To illustrate the use of the reconstruction formula (\ref{finalequation}) we apply it to the truncation described in subsection \ref{LPA}, the local potential approximation.
We apply the same type of truncation ansatz for the average and the bare action, with different potentials $U_{k,\Lambda}$ and $\check{U}_{\Lambda}$ though.
They are given by (\ref{actionLPA}) and
\begin{equation}\label{bareactionLPA}
S_{\Lambda}[\phi]=\vol\Big\{ \frac{1}{2}\partial_{\mu}\phi\partial^{\mu}\phi + \check{U}_{\Lambda}(\phi)\Big\}
\end{equation}
Inserting (\ref{actionLPA}) and (\ref{bareactionLPA}) into (\ref{finalequation}) the differential equation for the bare potential $\check{U}_{\Lambda}$ is easily
worked out:
\begin{equation}
U_{k,\Lambda}(\phi)-\check{U}_{\Lambda}(\phi)=v_{d}\int_{0}^{\Lambda^{2}}\!\!\textrm{d}y \;y^{d/2-1}\ln{ \Big\{\big[ y+\mathcal{R}_{k}(y) +{\check{U}''}_{\Lambda}(\phi) \big]M^{-2} \Big\}  }
\end{equation}
Inserting the optimized cutoff we obtain
\begin{eqnarray}\label{localpotential0}
U_{k,\Lambda}(\phi)-\check{U}_{\Lambda}(\phi)& = & 2v_{d}d^{-1}k^{d}\ln{\Big\{\big[ k^{2} + {\check{U}''}_{\Lambda}(\phi)\big]M^{-2}\Big\}} 
\nonumber\\
& & {} + v_{d}\int_{k^{2}}^{\Lambda^{2}}\!\!\textrm{d}y\; y^{d/2-1}\;\ln{ \Big\{\big[ y+{\check{U}''}_{\Lambda}(\phi) \big]M^{-2} \Big\}  }
\end{eqnarray}
In $d=4$ dimensions this yields explicitly, with $v_{4}=(32\pi^{2})^{-1}$,
\begin{eqnarray}\label{localpotential}
U_{k,\Lambda}(\phi) & = &\check{U}_{\Lambda}(\phi)+ \frac{1}{2}v_{4}\Big\{ \Lambda^{4}\ln{\Big[ \frac{\Lambda^{2}+{\check{U}''}_{\Lambda}(\phi) }{M^{2}}\Big]}
-\Big({\check{U}''}_{\Lambda}(\phi) \Big)^{2}\ln{\Big[ \frac{\Lambda^{2}+{\check{U}''}_{\Lambda}(\phi) }{k^{2}+{\check{U}''}_{\Lambda}(\phi)}\Big]}
\nonumber\\
& & {} -\frac{1}{2}(\Lambda^{4}-k^{4})+{\check{U}''}_{\Lambda}(\phi)(\Lambda^{2}-k^{2}) \Big\}
\end{eqnarray}

As a consistency check we can make a quartic ansatz for the local potential,
\begin{equation}\label{LPApoly}
{\check{U}}_{\Lambda}(\phi)=\frac{1}{2}\;\check{m}_{\Lambda}^{2}\;\phi^{2}+\frac{1}{4!}\;\check{\lambda}_{\Lambda}\;\phi^{4},
\end{equation}
insert it into (\ref{localpotential}), and let $k\rightarrow 0$. In this way we obtain the familiar formula for the standard 1-loop effective potential
$U_{0,\Lambda}\equiv U_{\Lambda}\equiv U_{\textrm{eff}}$ in presence of an UV cutoff. (The latter can be eliminated in the usual way after imposing
a renormalization condition.)

What we are actually interested in is the case $k=\Lambda$. Since, for the optimized cutoff, $\EA{\Lambda,\Lambda}=\EA{\Lambda}$ we may use
$U_{\Lambda, \Lambda}=U_{\Lambda}$ where $U_{\Lambda}\equiv U_{k=\Lambda}$ is the average potential appearing in the LPA ansatz for
solutions of the UV cutoff-free FRGE: $\EA{k}[\phi]=\vol \Big[  \tfrac{1}{2}(\partial\phi)^{2} + U_{k}(\phi)\Big]$. For this case (\ref{localpotential0}) boils down to
\begin{equation}\label{LPAdiffeq}
U_{\Lambda}(\phi)-\check{U}_{\Lambda}(\phi)=\frac{2v_{d}}{d}\;\Lambda^{d}\;\ln{\Big[  \frac{\Lambda^{2}+{\check{U}''}_{\Lambda}(\phi) }{M^{2}} \Big]}
\end{equation}
This, now, is an ordinary differential equation for the bare potential $\check{U}_{\Lambda}(\phi)$, albeit one of a somewhat unusual type.

In general the solution $\check{U}_{\Lambda}(\phi)$ of (\ref{LPAdiffeq}) for a given $U_{\Lambda}(\phi)$ will not have a simple
(polynomial, say) form. However, if we truncate $\check{U}_{\Lambda}(\phi)$  down to the polynomial (\ref{LPApoly}), and
make an analogous $\phi^2+\phi^4$ ansatz for $U_{\Lambda}(\phi)$, with coefficients $m_{\Lambda}^2$ and $\lambda_{\Lambda}$,
respectively,  the differential equation implies two algebraic equations relating the effective parameters
$(m_{\Lambda},\lambda_{\Lambda})$ to the bare ones, $(\check{m}_{\Lambda},\check{\lambda}_{\Lambda})$. They read, for
$d=4$,
\begin{equation}\label{LPAtruncation0}
m_{\Lambda}^2-\check{m}_{\Lambda}^2=\frac{\check{\lambda}_{\Lambda}}{64\pi^2}\frac{\Lambda^4}{\Lambda^2+\check{m}_{\Lambda}^2}
\end{equation}
\begin{equation}\label{LPAtruncation1}
\lambda_{\Lambda}-\check{\lambda}_{\Lambda}=-\frac{3\check{\lambda}_{\Lambda}^2}{64\pi^2}
\Big(\frac{\Lambda^2}{\Lambda^2+\check{m}_{\Lambda}^2}\Big)^2
\end{equation}
Since, at least in $d=4$, the scalar theory has no particularly interesting UV behavior we shall not study these 
relations any further here.


\section{Induced cosmological constant:\\
conceptual lessons from a toy model}\label{COSMO}

In this section we illustrate the relationship between the bare and the average action by means of a simple explicit example which
is also of interest in its own right: the cosmological constant induced by a (scalar, say) matter field quantized in a classical
gravitational background.

We assume that the scalar has no interactions except with the classical metric $g_{\mu\nu}$. 
Being interested in the induced 
cosmological constant  we retain the $\vol\sqrt{g}$ invariant in the bare and average action, respectively, but
discard terms involving derivatives of   $g_{\mu\nu}$: 
\begin{equation}\label{cosmobareaction}
S_{\Lambda}[\chi]=\frac{1}{2}\vol\sqrt{g}\Big[g^{\mu\nu}\partial_{\mu}\chi\partial_{\nu}\chi +\check{m}^2\chi^2\Big] +
\check{C}_{\Lambda}\vol\sqrt{g}
\end{equation}
\begin{equation}\label{cosmoeffaction}
\EA{k,\Lambda}[\phi]=\frac{1}{2}\vol\sqrt{g}\Big[g^{\mu\nu}\partial_{\mu}\phi\partial_{\nu}\phi +{m}^2\phi^2\Big] +
{C}_{k,\Lambda}\vol\sqrt{g}
\end{equation}
The solution $\EA{k}[\phi]$ of the $\Lambda$-free FRGE has a structure similar to (\ref{cosmoeffaction}) involving a
running parameter $C_k$. The three $C$-factors $\check{C}_{\Lambda}$, $C_{k,\Lambda}$ and $C_k$ are related to the corresponding
cosmological constants $\bar{\lambda}$ by $C\equiv (\bar{\lambda}/8\pi G)$
where Newton's constant $G$ does not run in the approximation considered. Furthermore, for the purposes of this demonstration, the running of the masses is also neglected.

\subsection{The flow equations for $C_{k,\Lambda}$ and $C_k$ }\label{floweqs}

The flow equation for $\EA{k,\Lambda}[\phi]$ is a slight generalization of (\ref{frge}) with the flat metric replaced by 
$g_{\mu\nu}$ everywhere. In particular, the operator $\hat{p}^2\equiv -D^2$ is now
to be interpreted as the Laplace-Beltrami operator constructed
with the metric $g_{\mu\nu}$. Upon inserting (\ref{cosmoeffaction}) the  FRGE assumes the form
\begin{equation}\label{cosmofrge}
k\partial_kC_{k,\Lambda}\vol\sqrt{g}=\frac{1}{2}\textrm{Tr}\Big[\theta(\Lambda^2+D^2)\;\mathcal{K}(-D^2)^{-1}\;
k\partial_k\mathcal{R}_k(-D^2)\Big]
\end{equation}
with $ \mathcal{K}(\hat{p}^2)\equiv \hat{p}^2+m^2+\mathcal{R}_k(\hat{p}^2)$.
To make eq.(\ref{cosmofrge}) consistent we may retain only the volume term $\propto$  $\vol\sqrt{g}$ in the derivative 
expansion of the trace on its RHS. It is easily found by inserting a flat metric, for instance:
\begin{equation}
k\partial_kC_{k,\Lambda}=\frac{1}{2}\int\!\!\frac{\textrm{d}^dp}{(2\pi)^d}\;\theta(\Lambda^2-p^2)
\;\frac{k\partial_k\mathcal{R}_k(p^2)}{ {p}^2+m^2+\mathcal{R}_k({p}^2)}
\end{equation}
Using the optimized cutoff this integral can be evaluated explicitly:
\begin{equation}\label{cosmofrge1}
k\partial_kC_{k,\Lambda}=\frac{4v_d}{d}\Big(\frac{k^2}{k^2+m^2}\Big)\;k^d
\end{equation}
We observe that the RHS of (\ref{cosmofrge1}) has become independent of the cutoff $\Lambda$. 

Inserting the 
$\EA{k}$-ansatz (involving $C_k$) into the $\Lambda$-free flow equation we find  eq.(\ref{cosmofrge1}), too,
 this time for $C_k$. 
Hence
$k\partial_kC_k=k\partial_kC_{k,\Lambda}$ for all $\Lambda\geq k$.

If $k\gg m$, eq.(\ref{cosmofrge1}) yields the familiar $k^d$-running of the cosmological constant; it is this scale 
dependence that would  result from summing up the zero point energies of the (massless) field modes. If $k\ll m$
 the running is much weaker since the RHS of (\ref{cosmofrge1}) contains a suppression factor $(k/m)^2\ll 1$. 
This is a typical decoupling phenomenon: In the regime $k\ll m$ the physical mass $m$ is the active IR cutoff.

The RG equation (\ref{cosmofrge1}) has the solution
\begin{equation}\label{cosmosol}
C_{k,\Lambda}= C_{\textrm{ren}}+\frac{2v_d}{d}\int^{k^2}_{0}\!\!\textrm{d}y\frac{y^{\frac{d}{2}}}{y+m^2}
\end{equation}
Here we fixed a specific RG trajectory by imposing the renormalization condition 
\protect{$C_{k=0,\Lambda\rightarrow\infty}=C_{\textrm{ren}}$} with $\bar{\lambda}_{\textrm{ren}}\equiv (8\pi G)C_{\textrm{ren}}$ the 
``renormalized cosmological constant", to be determined experimentally in principle. For $m=0$ in particular, since 
$C_k=C_{k,\Lambda}$ here,
\begin{equation}\label{cosmosolA}
C_k=C_{k,\Lambda}=C_{\textrm{ren}}+ 4d^{-2}\;v_d\; k^d
\end{equation}
If $d=4$, say, in standard notation,
\begin{equation}\label{cosmosolB}
\bar{\lambda}_{k}=\bar{\lambda}_{\textrm{ren}}+\frac{1}{16\pi^2}\;G_0\; k^4
\end{equation}
The scalar being massless, this running of  effective cosmological constant has the same structure as in
pure quantum gravity \cite{mr}.

\subsection{Exact forms of $W_{k,\Lambda}$ and $\EA{k,\Lambda}$}\label{exactforms}

Since $S_{\Lambda}$ is quadratic in $\chi$ the functional integral (\ref{wkl}) for $W_{k,\Lambda}[J]$, appropriately generalized to a curved background, can be solved exactly:
\begin{eqnarray}
W_{k,\Lambda}[J] &=& \frac{1}{2}\vol\sqrt{g}\; J\; \big[  -D^2+ \check{m}^2+\mathcal{R}_k(-D^2)\big]^{-1} J-
\check{C}_{\Lambda}\vol\sqrt{g}-
\nonumber\\
& & {}-\frac{1}{2}\textrm{Tr}_{\Lambda}\ln{\Big[\Big( -D^2+ \check{m}^2+\mathcal{R}_k(-D^2)\Big)M^{-2}\Big]}
\end{eqnarray}
In this simple case we can compute $\EA{k,\Lambda}$ directly from the very definition of the EAA, eq.(\ref{effectiveaction}):
\begin{eqnarray}\label{cosmoeffaction2}
\EA{k,\Lambda}[\phi] &=& \frac{1}{2}\vol\sqrt{g}\Big(g^{\mu\nu}\partial_{\mu}\phi\partial_{\nu}\phi +\check{m}^2\phi^2 \Big)+
\check{C}_{\Lambda}\vol\sqrt{g}+
\nonumber\\
& & {} + \frac{1}{2}\textrm{Tr}_{\Lambda}\ln{\Big[\Big( -D^2+ \check{m}^2+\mathcal{R}_k(-D^2)\Big)M^{-2}\Big]}
\end{eqnarray}

\subsection{The difference between the bare $\check{C}_{\Lambda}$ and \\
the effective $C_{\Lambda,\Lambda}$}
\label{differences}
By performing a derivative expansion of $\textrm{Tr}_{\Lambda}\ln{[\cdots]}$ in (\ref{cosmoeffaction2}) we obtain the
scalar's contribution to the induced cosmological constant ($\int\sqrt{g}$ term), the induced Newton constant ($\int\sqrt{g}R$
term), and similarly to the higher derivative terms. Here we are interested in the cosmological constant only, and comparing
(\ref{cosmoeffaction2}) to (\ref{cosmoeffaction}) yields
\begin{eqnarray}\label{cosmodiff}
C_{k,\Lambda}-\check{C}_{\Lambda} & = &\frac{1}{2} \Big[\vol\sqrt{g}\Big]^{-1}\;\textrm{Tr}_{\Lambda}\ln{\Big[\cdots\Big]}\Big|_{\int\sqrt{g} \textrm{ term}}
\nonumber\\
& & {} = \frac{1}{2}\int\!\!\frac{\textrm{d}^dp}{(2\pi)^d}\;\theta(\Lambda^2-p^2)\;\ln{\Big(\big[p^2+m^2 +\mathcal{R}_k(p^2)\big]M^{-2}\Big)}
\end{eqnarray}
Employing the optimized cutoff again, (\ref{cosmodiff}) evaluates to
\begin{equation}\label{cosmodiff2}
C_{k,\Lambda}=\check{C}_{\Lambda} +\frac{2v_d}{d}\; k^d\;\ln{\Big(\frac{k^2+m^2}{M^2}\Big)}
+v_d\int_{k^2}^{\Lambda^2}\!\!\textrm{d}y \;y^{d/2-1}\ln{\Big(\frac{y^2+m^2}{M^2}\Big)}
\end{equation}
Note that in (\ref{cosmodiff}) and (\ref{cosmodiff2}) we replaced $\check{m}$ with $m$ since comparing the $\phi^2$-terms in 
(\ref{cosmoeffaction2}) and (\ref{cosmoeffaction}), respectively, implies that $\check{m}=m$ within the simple truncation used. 

For $m=0$ and $d=4$, say, eq.(\ref{cosmodiff2}) implies the following explicit result for the running effective cosmological
constant in terms of the bare one:
\begin{equation}
C_{k,\Lambda}=\check{C}_{\Lambda}+v_4\Big[  \Lambda^4\ln{(\Lambda/M)}-\frac{1}{4}(\Lambda^4-k^4)\Big]
\end{equation}
Taking the $k$-derivative of the function $C_{k,\Lambda}$ in (\ref{cosmodiff2}), at fixed $\Lambda$, we see that it does indeed 
satisfy the flow equation (\ref{frge1}).

For arbitrary $d$ and $m$, the limit $k\rightarrow\Lambda$ of eq.(\ref{cosmodiff2}) reads
\begin{equation}\label{barecc0}
\check{C}_{\Lambda}=C_{\Lambda,\Lambda}-\frac{2v_d}{d}\;\Lambda^d\;\ln{\Big(\frac{\Lambda^2+m^2}{M^2}\Big)}
\end{equation}
This equation tells us how, for a given effective cosmological constant $C_{\Lambda,\Lambda}$, the bare one, 
$\check{C}_{\Lambda}$, must be adjusted in order to give rise to the prescribed effective one. The value
of $C_{\Lambda,\Lambda}$ in turn depends on the RG trajectory chosen, i.e., in this simple situation, on the value of 
$C_{\textrm{ren}}$. In fact, from the explicit solution (\ref{cosmosol}) we get
\begin{equation}\label{effcc0}
C_{\Lambda,\Lambda}= C_{\textrm{ren}}+\frac{2v_d}{d}\int^{\Lambda^2}_{0}\!\!\textrm{d}y\frac{y^{\frac{d}{2}}}{y+m^2}
\end{equation}

\subsection{Some general lessons}\label{conceptual}

The above simple formulae illustrate various conceptual lessons of general significance.

\noindent
{\bf{(A) Nonuniqueness of the bare action.}} Let us consider the massless case $m=0$ which may serve as a toy model 
for gauge fields. Then the cosmological constant in the bare action is
\begin{equation}\label{barecc}
\check{C}_{\Lambda}=C_{\Lambda,\Lambda}-4d^{-1}v_d\;\Lambda^d\;\ln{(\Lambda/M)}
\end{equation}
while the one in $\EA{k,\Lambda}$ and $\EA{k}$ at $k=\Lambda$ reads
\begin{equation}\label{effcc}
C_{\Lambda,\Lambda}= C_{\textrm{ren}}+ 4d^{-2}v_d\;\Lambda^d=C_{k=\Lambda}
\end{equation}
Recall that the mass parameter $M$ was introduced in (\ref{measure}) in order to make $\mathcal{D}_{\Lambda}\chi$ dimensionless.
How should this parameter be chosen? Our choice for $M$ will affect the bare cosmological constant (\ref{barecc})
but not the effective one, eq.(\ref{effcc}). The {\it{effective}} cosmological constant $C_{k=\Lambda}$ will always be
proportional to $\Lambda^d$ for $\Lambda\rightarrow\infty$ and approach {\it{plus}} infinity.

As a first choice consider $M=\textrm{const}$, i.e. $M$ is a positive constant independent of $\Lambda$. Then, according to
(\ref{barecc}), the {\it{bare}} cosmological constant $\check{C}_{\Lambda}$ is proportional to $-\Lambda^d\ln{\Lambda}$ for
$\Lambda\gg M$ and it approaches {\it{minus}} infinity in the limit $\Lambda\rightarrow\infty$.

As a second choice assume $M$ is proportional to the UV cutoff, $M=c\Lambda$, with some constant $c>0$. Then 
$\check{C}_{\Lambda}=C_{\textrm{ren}}+ 4d^{-2}v_d\Lambda^d\{1-d\ln{c}\}$ diverges proportional to $\Lambda^d$ if 
$c\neq\exp{(1/d)}$, and depending on the value of $c$ it might approach  $-\infty$ or $+\infty$.
In the special case $c=\exp{(1/d)}$ the bare cosmological constant $\check{C}_{\Lambda} $ equals $C_{\textrm{ren}}$ for all
$\Lambda$, i.e. it is {\it{finite}} even in the limit $\Lambda\rightarrow\infty$. Also $c=1$ is special: in this case,
accidentally, the bare and the effective average action contain {\it{the same}} cosmological constant: 
$\check{C}_{\Lambda}=C_{\Lambda,\Lambda}$.

Even though they can lead to dramatically different bare actions, the various choices for $M$ are all physically 
equivalent.\footnote{The case $M=$const is mentioned here only to give an extreme example.
Clearly it is quite unnatural. In a standard discussion of the continuum limit one would express the field
in UV cutoff units which amounts to $M=\Lambda $. (In presence of the second cutoff $k$ also other
options could be convenient.)}
The ordinary effective action  and the EAA are independent of $M$. Changing $M$ simply
amounts to shifting contributions from the measure into the bare action or vice versa.

This illustrates a general lesson which, while true everywhere in quantum field theory, is particularly important in the 
asymptotic safety context: It makes no sense to talk about a bare action unless one has specified a measure before;
neither $\mathcal{D}_{\Lambda}\chi$ nor $\exp{[-S_{\Lambda}]}$ have a physical meaning separately, only the combination
$\int\mathcal{D}_{\Lambda}\chi\exp{[-S_{\Lambda}]}$ has. Here we illustrated this phenomenon by a simple rescaling
of the integration variable but clearly it extends to more general transformations of $\chi$ whose Jacobian is interpreted
as changing the action $S_{\Lambda}$ to a new one, $S^{'}_{\Lambda}$.

The concrete lesson for the asymptotic safety program is that one should not expect a fixed point solution of the FRGE,
$\Gamma_*$, to correspond to a unique bare action.

\noindent
{\bf{(B) Flow equation for $\mathbf{S_{\Lambda}}$?} } Our general strategy is to first  solve the FRGE for the EAA. Then,
 having found a concrete RG trajectory 
$k\mapsto \EA{k}$, we determine which trajectory of bare actions $\Lambda\mapsto S_{\Lambda}$ gives
rise to this average action.
It is therefore natural to ask if there is a flow  equation that governs the $\Lambda$-dependence of the bare 
actions defined in this way. 

Using the above formulae we can easily  answer this question for the cosmological constant term in $S_{\Lambda}$.
If we take  the $\Lambda$-derivative of (\ref{barecc0}) or (\ref{effcc0}) and exploit that
\begin{equation}
\Lambda\partial_{\Lambda}C_{\Lambda,\Lambda}=\frac{4v_d}{d}\;\frac{\Lambda^{d+2}}{\Lambda^2+m^2}
\end{equation}
which obtains by differentiating (\ref{effcc0}), we find 
\begin{equation}\label{cosmoRGeq}
\Lambda\partial_{\Lambda}\check{C}_{\Lambda}=-\frac{4v_d}{d}\;\Lambda^d\;\Big[\frac{d}{2}\ln{\Big(\frac{\Lambda^2+m^2}{M^2}\Big)}     
-\frac{\Lambda\partial_{\Lambda}M}{M}\Big]
\end{equation}
This equation tells how the bare action must change when $\Lambda$ is sent to infinity, given the requirement that the
parameter $C_{k=0}$ in the ordinary effective action assumes the prescribed value $C_{\textrm{ren}}$. Obviously
{\it{the RG equation for the bare cosmological constant
 is quite different from the corresponding equation at the level of the effective average action}},
eq.(\ref{cosmofrge1}). 

So, for constructing a path integral describing an asymptotically safe theory, why not use a full fledged functional flow
equation for the bare action? Why is the RG flow of $\EA{k}$  crucial for the QEG program, while $S_{\Lambda}$ plays only
a secondary role? There are at least two answers to these questions:

\noindent
{\bf{(i) Absence of divergences in observable quantities.}} As we already briefly mentioned {\it{the property of asymptotic safety 
is decided about at the
effective rather than bare level.}} By its very definition, asymptotic safety requires observable quantities
such as scattering cross sections to be free from  divergences. Since the $S$-matrix elements are 
essentially functional derivatives of $\Gamma\equiv \Gamma_{k=0}$ this requires the ordinary effective
action to be free from such divergences. This is indeed the case if $\Gamma$ is connected to a UV fixed point 
$\Gamma_*$ by a regular RG trajectory. So, in order to test wether this condition is satisfied we need to 
know the $\Gamma_k$-flow. The concomitant $S_{\Lambda}$-flow is of no direct physical relevance.
In principle is is even conceivable that, while $\Gamma_k$ approaches to a fixed point  in the UV, the bare action
does not; the resulting theory could nevertheless have completely acceptable physical properties.
(Below we shall encounter a simple, albeit somewhat artificial example where this happens.) 

For these reasons the basic tool in searching for asymptotic safety is the flow equation for the EAA
and not its analog for the bare action.

\noindent
{\bf{(ii) Effective field theory properties.}} 
 We would like the scale dependent functional obtained by solving the flow equation 
to have a chance of defining an effective field theory in the sense that its tree level evaluation at some scale 
approximately describes all quantum effects with this typical scale. For $\EA{k}$ this is indeed the 
case\footnote{Of course we are not saying here that $\EA{k}$ necessarily provides a numerically precise description. To what degree
this is actually possible (fluctuations are small, etc.) depends on the details of the physical situation.}, 
but not for $S_{\Lambda}$. The reason is that, given $S_{\Lambda}$, there is still a functional integration to be performed in 
order to go over to the effective level; using $\EA{k}$ instead, it has been performed already.

The above toy model illustrates this point: From eq.(\ref{cosmosolA}) or eq.(\ref{cosmosolB}) we conclude that for every finite
$\bar{\lambda}_{\textrm{ren}}\equiv (8\pi G)C_{\textrm{ren}}$ the {\it{running effective}} cosmological constant
$\bar{\lambda}_k\equiv (8\pi G)C_{k}$ becomes large and positive for growing $k$ and finally approaches {\it{plus}} infinity
for $k\rightarrow\infty$. Applying the effective field theory interpretation we would  insert this $\bar{\lambda}_k$ into the effective Einstein equation. It then 
predicts that, at high momentum scales, spacetime  is strongly curved and has {\it{positive}} curvature. 

From the above remarks  it is clear that the {\it{running bare}} action does not contain this information.
Depending on our choice for $M$ the bare cosmological constant $\check{C}_{\Lambda}$ approaches to $+\infty$,$-\infty$ or a finite value where $\Lambda\rightarrow\infty$. So clearly it would not make any sense to insert it
into Einstein's equation in order to ``RG improve" it.

\section{ QEG and the Einstein-Hilbert truncation}\label{QEGtruncation}

In this section we apply the strategies developed above to QEG. We generalize the construction of the gravitational average
action by introducing an UV cutoff, and we determine the resulting regularized bare action in terms of $\Gamma_k$. 

In this section we use the same notations and conventions as in reference \cite{mr} to which the reader is referred 
for further details.

\subsection{Covariant UV cutoff in the background approach}\label{covariantUV} 

The construction of the gravitational average actions starts out from a path integral 
$\int\mathcal{D}\gamma_{\mu\nu}\exp{(-S[\gamma_{\mu\nu}])}$.
First we introduce a background metric $\bar{g}_{\mu\nu}(x)$, decompose the integration variable as 
$\gamma_{\mu\nu}\equiv \bar{g}_{\mu\nu}+ h_{\mu\nu}$, and gauge-fix the resulting path integral over $h_{\mu\nu}$.
It is this integral that we make well defined by introducing an UV cutoff into the measure along with an IR-suppression
term $\Delta_kS$ analogous to (\ref{IR}):
\begin{equation}\label{qegpath}
\int\mathcal{D}_{\Lambda}h\;\mathcal{D}_{\Lambda}C\;\mathcal{D}_{\Lambda}\bar{C}\;\exp{\Big(-\widetilde{S}_{\Lambda}[h,C,\bar{C};\bar{g}]
-\Delta_kS[h,C,\bar{C};\bar{g}]\Big)}
\end{equation} 
Here $C^{\mu}$ and $\bar{C}_{\mu}$ are the Fadeev-Popov ghosts, and the total bare action,
$
\widetilde{S}_{\Lambda}\equiv\BA{\Lambda}+\BA{\textrm{gf},\Lambda}+\BA{\textrm{gh},\Lambda}
$,
which is allowed to depend on $\Lambda$, includes the gauge fixing term $\BA{\textrm{gf},\Lambda}$ and the ghost
action $\BA{\textrm{gh},\Lambda}$.

The new feature in (\ref{qegpath}) is the UV regularized measure. It is defined as follows. We start by using the (fixed, but not
concretely specified) background metric $\bar{g}_{\mu\nu}(x)$ in order to construct the covariant Laplacians
$\bar{D}^2\equiv \bar{g}^{\mu\nu}\bar{D}_{\mu}\bar{D}_{\nu}$ , appropriate for symmetric second  rank tensor,
vector, and co-vector fields, respectively. Then, at least in principle, we determine a complete set of eigenmodes $\{u^{\kappa m}(x)\}$ of
these Laplacians and expand the integration variables $h_{\mu\nu}(x)$, $C^{\mu}(x)$, and $\bar{C}_{\mu}(x)$ in terms of those.
For instance, 
$
h_{\mu\nu}(x)=\sum_{\kappa,m}h_{\kappa m}\;u^{km}_{\mu\nu}(x),
$
and similarly for the ghosts. Here $\kappa$ denotes the negative eigenvalue, $\bar{D}^2u^{\kappa m}=-\kappa u^{\kappa m}$,
and $m$ is a degeneracy index. We implement the UV cutoff by restricting the expansion to eigenfunctions with eigenvalues 
$\kappa$ smaller than a given $\Lambda^2$.
Hence the measure reads in analogy with (\ref{measure})
\begin{equation}\label{qegmeasure}
\int\mathcal{D}_{\Lambda}h=\prod_{\kappa\in[0,\Lambda^2]}\prod_m\int_{-\infty}^{\infty}\!\!\textrm{d}h_{\kappa m}
\;M^{-[h_{\kappa m}]}
\end{equation}
and likewise for the ghosts.

The remaining steps in the construction of the gravitational average action, now denoted
$\EA{k,\Lambda}[\bar{h}_{\mu\nu},\xi^{\mu},\bar{\xi}_{\mu};\bar{g}_{\mu\nu}]$, with the expectation values
$\bar{h}_{\mu\nu}\equiv\langle {h}_{\mu\nu}\rangle$, $\xi^{\mu}\equiv\langle C^{\mu}\rangle$ and 
$\bar{\xi}_{\mu}\equiv\langle\bar{C}_{\mu}\rangle$, proceed exactly  as in \cite{mr}
(coupling to sources, Legendre transformation, subtraction of $\Delta_kS$ at the classical level).

The key properties of the functional thus defined are the exact FRGE and the integro-differential equation which it satisfies.
The flow equation reads
\begin{equation}\label{qegfloweq}
k\partial_k\EA{k,\Lambda}[\bar{h},\xi,\bar{\xi};\bar{g}]=\frac{1}{2}\textrm{STr}_{\Lambda}\Big[\Big(\EA{k,\Lambda}^{(2)}+\widehat{\mathcal{R}}_{k}\Big)^{-1}k\partial_{k}\widehat{\mathcal{R}}_{k}\Big]
\end{equation}
Here the supertrace ``STr" implies the extra minus sign in the ghost sector. In fact, the cutoff operator $\widehat{\mathcal{R}}_{k}$ and the Hessian $\EA{k,\Lambda}^{(2)}$ are matrices in the space of dynamical fields $\bar{h},\xi$ and $\bar{\xi}$.
The background covariant regularization of the measure entails the appearance of the restricted trace
\begin{equation}\label{qegtrace}
\textrm{STr}_{\Lambda}[\cdots]\equiv\textrm{STr}\Big[\theta(\Lambda^2+\bar{D}^2)[\cdots]\Big]
\end{equation}
Note that in this construction the background metric $\bar{g}_{\mu\nu}(x)$ is crucial not only for the gauge fixing and the 
IR cutoff, but also for implementing the UV cutoff.

In parallel with (\ref{qegfloweq}) we shall also consider the usual FRGE of QEG without an UV cutoff. Its solutions will
be denoted $\EA{k}[\bar{h},\xi,\bar{\xi};\bar{g}]$. The discussion of the relation between $\EA{k,\Lambda}$ and $\EA{k}$
parallels the one in Subsection \ref{section4} above. In particular, if we use the optimized cutoff, then eq.(\ref{inicond1})
holds true in QEG, too:
\begin{equation}\label{qeginicond}
\EA{\Lambda,\Lambda}[\bar{h},\xi,\bar{\xi};\bar{g}]=\EA{\Lambda}[{h},\xi,\bar{\xi};\bar{g}]
\end{equation}

The integro-differential equation analogous to (\ref{integroeq0}) reads in QEG\footnote{Except for the UV cutoff, 
eq.(\ref{qeginteq}) coincides with eq.(2.34) in \cite{mr} for vanishing BRS sources $\beta$ and $\tau$ which we do not
need here.}

\begin{align}\label{qeginteq}
\exp{\Big(-\EA{k,\Lambda}[\bar{h},\xi,\bar{\xi};\bar{g}]\Big)}= &
\int\mathcal{D}_{\Lambda}h\mathcal{D}_{\Lambda}C\mathcal{D}_{\Lambda}\bar{C}
\exp{\Big[
-\widetilde{S}_{\Lambda}[h,C,\bar{C};\bar{g}]}-
\nonumber\\ 
- & \Delta_k S[h-\bar{h},C-\xi,\bar{C}-\bar{\xi};\bar{g}] 
+\vol({h}_{\mu\nu}-\bar{h}_{\mu\nu})\frac{\delta\EA{k,\Lambda}}{\delta\bar{h}_{\mu\nu}}
\nonumber\\ 
+ & \vol(C^{\mu}-\xi^{\mu})\frac{\delta\EA{k,\Lambda}}{\delta\xi^{\mu}}
+\vol(\bar{C}_{\mu}-\bar{\xi}_{\mu})\frac{\delta\EA{k,\Lambda}}{\delta\bar{\xi}^{\mu}}\Big]
\end{align}
An important property of $\EA{k,\Lambda}[\bar{h},\xi,\bar{\xi};\bar{g}]$ is its invariance under background gauge transformations.
Under these transformations all four arguments, $\bar{h},\xi,\bar{\xi}$ and $\bar{g}$, transform as tensors of the 
corresponding rank. This property is preserved by the specific UV regularization we have chosen.

An alternative notation for the average action is $\EA{k,\Lambda}[g,\bar{g},\xi,\bar{\xi}]\equiv 
\EA{k,\Lambda}[\bar{h},\xi,\bar{\xi};\bar{g}]$
where ${g}_{\mu\nu}\equiv \bar{g}_{\mu\nu} +\bar{h}_{\mu\nu} $ denotes the complete classical metric, the expectation value of 
$\gamma_{\mu\nu}$.

\subsection{The bare action at one loop}\label{oneloop}

As in the scalar case above, we would like to use the information contained in  a given solution
$\EA{k}[\bar{h},\xi,\bar{\xi};\bar{g}]$ of the $\Lambda$-free FRGE in order to find out which $\Lambda$-dependence must
be given to the (total) bare action $\widetilde{S}_{\Lambda}$ if we want the path integral to possess a well defined limit
$\Lambda\rightarrow\infty$ and to reproduce the prescribed $\EA{k}$. The key relations are eq.(\ref{qeginicond}) which allows us
to interpret the known $\EA{k=\Lambda}$ as  $\EA{\Lambda,\Lambda}$, and the integro-differential equation (\ref{qeginteq}).
When evaluated at $k=\Lambda\rightarrow\infty$ the latter yields the sought for relationship between the bare and the average
action. If we restrict ourselves to the 1-loop level, its derivation proceeds as in Subsection \ref{sectionsaddle}, with
the result
\begin{equation}\label{qegbareeff}
\EA{\Lambda,\Lambda}[\bar{h},\xi,\bar{\xi};\bar{g}]=\widetilde{S}_{\Lambda}[\bar{h},\xi,\bar{\xi};\bar{g}]
+\frac{1}{2}\textrm{STr}_{\Lambda}\ln{\Big[\Big(\widetilde{S}_{\Lambda}^{(2)}+\widehat{\mathcal{R}}_{\Lambda}\Big)[\bar{h},\xi,\bar{\xi};\bar{g}]\;\mathcal{N}^{-1}\Big]}
\end{equation}
Here $\mathcal{N}$ is a block diagonal normalization matrix, equal to $M^d$ and $M^2$ in the graviton and the ghost sector,
respectively. For a given $\EA{\Lambda,\Lambda}$, eq.(\ref{qegbareeff}) is to be regarded a differential equation
for the complete bare action {\it{which includes all gauge fixing and ghost terms}}.

At first sight it might seem puzzling that the formalism  itself tells us here which gauge fixing is to be used.
This puzzle is resolved, however, by recalling a general feature of the background gauge fixing technique \cite{back} used here:
While   $\EA{k,\Lambda}$ is a gauge (i.e. diffeomorphism) invariant functional of its arguments, it is {\it{not}} 
independent of the (background-type, but otherwise arbitrary) gauge fixing condition.
Therefore $\EA{\Lambda,\Lambda}$ does indeed contain information about the gauge fixing condition.

\subsection{The twofold Einstein-Hilbert truncation}\label{EHtruncation}

The explicit computation of the bare action is difficult, even at the 1-loop level of eq.(\ref{qegbareeff}). In practice
one has to truncate the space the actions $\EA{k}$ and $\widetilde{S}_{\Lambda}$ ``live" in. Here we are going to analyze the 
simplest possibility, the Einstein-Hilbert truncation for both the effective and the bare action. As in \cite{mr}
we make the ansatz
\begin{eqnarray}\label{EHansatz}
\EA{k}[g,\bar{g},\xi,\bar{\xi}] & = &-(16\pi G_k)^{-1}\vol\sqrt{g}\Big(R(g)-2\bar{\lambda}_k\Big) 
+S_{\textrm{gh}}[g-\bar{g},\xi,\bar{\xi};\bar{g}]
\nonumber\\
& & {} + (32\pi G_k)^{-1}\vol\sqrt{\bar{g}}\bar{g}^{\mu\nu}(\mathcal{F}_{\mu}^{\alpha\beta}g_{\alpha\beta})\!
(\mathcal{F}_{\nu}^{\rho\sigma}g_{\rho\sigma})
\end{eqnarray}
The third term on the RHS of eq.(\ref{EHansatz}) 
is the gauge fixing term\footnote{We employ a non-dynamical gauge fixing parameter $\alpha=1$ here.} corresponding to the harmonic coordinate
condition, involving $\mathcal{F}_{\mu}^{\alpha\beta}\equiv \delta^{\beta}_{\mu}\bar{g}^{\alpha\gamma}\bar{D}_{\gamma}
-\tfrac{1}{2}\bar{g}^{\alpha\beta}\bar{D}_{\mu}$, and the second term is the associated ghost action. 
We make an analogous ansatz for the bare action:
\begin{eqnarray}\label{EHbareansatz}
\widetilde{S}_{\Lambda}[g,\bar{g},\xi,\bar{\xi}] & = &-(16\pi \check{G}_{\Lambda})^{-1}\vol\sqrt{g}\Big(R(g)-2\check{\bar{\lambda}}_{\Lambda}\Big) 
+S_{\textrm{gh}}[g-\bar{g},\xi,\bar{\xi};\bar{g}]
\nonumber\\
& & {} + (32\pi \check{G}_{\Lambda})^{-1}\vol\sqrt{\bar{g}}\bar{g}^{\mu\nu}(\mathcal{F}_{\mu}^{\alpha\beta}g_{\alpha\beta})\!
(\mathcal{F}_{\nu}^{\rho\sigma}g_{\rho\sigma})
\end{eqnarray}
Eq.(\ref{EHansatz}) contains the running dimensionful parameters $G_k$ and $\bar{\lambda}_k$. The corresponding
bare Newton and cosmological constant, respectively, are denoted $\check{G}_{\Lambda}$ and $\check{\bar{\lambda}}_{\Lambda}$.

We shall now insert $\EA{\Lambda,\Lambda}=\EA{k=\Lambda}$ with $\EA{k}$ given by (\ref{EHansatz}) into  
(\ref{qegbareeff}), along with (\ref{EHbareansatz}), and equate the coefficients of $\vol\sqrt{g}$ and $\vol\sqrt{g}R(g)$ on both
sides of the resulting equation. For this purpose it is sufficient to evaluate the traces for $\xi=\bar{\xi}=0$ and
$\bar{g}=g$ since the gauge fixing and ghost terms which are set to zero in this way do not contain independent information in this
truncation. (The Hessian $\tilde{S}_{\Lambda}^{(2)}$ is computed {\it{before}} setting $\xi=\bar{\xi}=\bar{h}=0$, of course.) The 
super trace has a derivative expansion of the form
\begin{equation}\label{qegDE}
\frac{1}{2}\textrm{STr}_{\Lambda}\ln{\Big[\Big(\widetilde{S}_{\Lambda}^{(2)}+\widehat{\mathcal{R}}_{\Lambda}\Big)[0,0,0;\bar{g}]\mathcal{N}^{-1}\Big]}\!=\!B_0\Lambda^d\vol\sqrt{g}+B_1\Lambda^{d-2}\vol\sqrt{g}R(g)+\cdots
\end{equation}
with dimensionless coefficients $B_0$ and $B_1$, respectively. Using (\ref{qegDE}) in (\ref{qegbareeff}) and equating 
the coefficients of the independent invariants we obtain two equations relating the effective to the bare parameters:
\begin{align}\label{Eqcoup1}
\frac{1}{G_{\Lambda}}-\frac{1}{\check{G}_{\Lambda}} =  -16\pi\; B_1\;\Lambda^{d-2},\qquad &
\frac{\bar{\lambda}_{\Lambda}}{G_{\Lambda}}-\frac{\check{\bar{\lambda}}_{\Lambda}}{\check{G}_{\Lambda}}
= 8\pi\; B_0\;\Lambda^d
\end{align}
It is convenient to introduce the dimensionless couplings 
\begin{align}\label{qegcoupling1}
g_{\Lambda} & \equiv  \Lambda^{d-2}G_{\Lambda} &
\check{g}_{\Lambda}  \equiv  \Lambda^{d-2}\check{G}_{\Lambda}
\nonumber\\
\lambda_{\Lambda} &\equiv  \Lambda^{-2}\bar{\lambda}_{\lambda} &
\check{\lambda}_{\Lambda} \equiv  \Lambda^{-2}\check{\bar{\lambda}}_{\Lambda}
\end{align}
As a result, the relations (\ref{Eqcoup1}) assume the form
\begin{subequations}\label{Eqcoup2}
\begin{align}
\frac{1}{g_{\Lambda}}-\frac{1}{\check{g}_{\Lambda}} & =  -16\pi \;B_1
\label{second}\\
\frac{{\lambda}_{\Lambda}}{g_{\Lambda}}-\frac{\check{\lambda}}{\check{g}_{\Lambda}}
& =  8\pi\; B_0
\label{third}
\end{align}
\end{subequations}
The equations \eqref{Eqcoup2} should allow  us to determine $\check{g}_{\Lambda}$ and $\check{\lambda}$ for
given $g_{\Lambda}$ and ${\lambda}_{\Lambda}$.

What remains to be done is the computation of $B_0$ and $B_1$. It is sketched in Appendix \ref{appendix}  where the results for an arbitrary  
dimension $d$ and the ``optimized" $\mathcal{R}_k$ are given in  eqs.(\ref{abb})-(\ref{aqq}). 
Since the conceptual issues we are interested in are 
the same in all dimensions, we set $d=4$ from now on. In this case,

\begin{subequations}\label{Bs}
\begin{align}
B_0 &= \frac{1}{32\pi^2}\Big[5\ln{(1-2\check{\lambda}_{\Lambda})}-5\ln{(\check{g}_{\Lambda})}  +Q_{\Lambda} \Big]
\label{b0}\\
B_1 &=\frac{1}{3}B_0+\Delta B_1
\label{b1}\\
\Delta B_1 &\equiv \frac{1}{16\pi^2}\frac{2-\check{\lambda}_{\Lambda}}{1-2\check{\lambda}_{\Lambda}}
\label{db}\\
Q_{\Lambda}&\equiv 12\ln{\big(\Lambda/M\big)}+b_0
\label{q}
\end{align}
\end{subequations}
with the constant $b_0\equiv -5\ln{(32\pi)}-\ln{2}$.

Thus, with (\ref{Bs}) the system of equations (\ref{Eqcoup2}) is known in explicit form. Actually, rather than working with 
 \eqref{second} and \eqref{third} as the two independent equations it is more convenient to use 
\eqref{third} together with the special linear combination of the two from which the logarithm drops out. Then we are 
left with
\begin{subequations}\label{qeggl}
\begin{align}
\frac{1}{\check{g}}\big(3+2\check{\lambda}\big) -& \frac{1}{g}\big(3+2\lambda\big) = 
\frac{3}{\pi}\;\frac{2-\check{\lambda}}{1-2\check{\lambda}}
\label{qegg}\\
\frac{\lambda}{g}-\frac{\check{\lambda}}{\check{g}} &=
\frac{1}{4\pi}\Big[5\ln{(1-2\check{\lambda})}-5\ln{\check{g}}  +Q \Big]
\label{qegl}
\end{align}
\end{subequations}
In writing down these relations we suppressed the subscript ``$\Lambda$".

\subsection{The map $(g,\lambda)\mapsto (\check{g},\check{\lambda})$}\label{5.3}

Unfortunately it is impossible to solve the system \eqref{qegg},\eqref{qegl} analytically
for the bare parameters. 
The best we can do is to solve (\ref{qegg}) for
$\check{g}:$ 
\begin{equation}\label{gcheck}
\check{g}=\big(3+2\check{\lambda}\big)\big(1-2\check{\lambda}\big)
\Big[ \frac{1}{g}\big(3+2\lambda\big)\big(1-2\check{\lambda}\big)
+ \frac{3}{\pi}\;\big(2-\check{\lambda}\big)\Big]^{-1}
\end{equation}
Inserting this expression  into \eqref{qegl} we obtain
an explicitly $\Lambda$-dependent transcendental equation for the bare cosmological constant:
$\check{\lambda}=\check{\lambda}(g,\lambda;\Lambda)$.


Conversely, it is straightforward to solve \eqref{qeggl} for the effective parameters as a function of 
the bare ones:
\begin{align}\label{5.101}
g(\check{g},\check{\lambda})& =\frac{3}{f_1(\check{g},\check{\lambda})-2f_2(\check{g},\check{\lambda})}&
\lambda(\check{g},\check{\lambda}) &=\frac{3f_2(\check{g},\check{\lambda})}{f_1(\check{g},\check{\lambda})
-2f_2(\check{g},\check{\lambda})}
\end{align}
The functions $f_1$ and $f_2$ are defined by:
\begin{align}
f_1(\check{g},\check{\lambda}) &\equiv \frac{3+2\check{\lambda}}{\check{g}}
-\frac{3}{\pi}\;\frac{(2-\check{\lambda})}{1-2\check{\lambda}}
\label{5.102}\\
f_2(\check{g},\check{\lambda}) &\equiv \frac{\check{\lambda}}{\check{g}}
+\frac{1}{4\pi}\Big[5\ln{\Big(\frac{1-2\check{\lambda}}{\check{g}}\Big)} +Q  \Big]
\label{5.103}
\end{align}
The physically relevant part of the effective coupling constant space is to the left of the boundary line 
$\lambda=1/2$ on which the $\beta$-functions diverge. The condition $\lambda<1/2$ requires 
$\check{\lambda}<1/2$ on the bare side. Because of the logarithm in \eqref{5.103} only positive values of
 the bare Newton constant are possible therefore, $\check{g}>0$.

\noindent
{\bf{(A) Fixing the parameter M.}} The map  $(g,\lambda)\mapsto (\check{g},\check{\lambda})$
given by the above equations, because of the parameter $Q\equiv Q_{\Lambda}\equiv 12\ln{(\Lambda/M)}+b_0$, 
is explicitly $\Lambda$-dependent; it defines a RG-time dependent diffeomorphism on some part of
$R^2$. This $\Lambda$-dependence can be removed by including appropriate factors of the
UV cutoff into the measure. If we set $M=c\Lambda$ with an arbitrary $c>0$ the quantity 
\begin{equation}\label{5.104a}
Q=12\ln{c} +b_0
\end{equation}
becomes a $\Lambda$-independent constant. Henceforth we shall adopt this choice. As a result, the map 
$(g,\lambda)\mapsto (\check{g},\check{\lambda})$ has no explicit dependence on any (UV or IR) cutoff.

Having no explicit cutoff dependence, $(g,\lambda)\mapsto (\check{g},\check{\lambda})$ maps ``effective''
RG trajectories $(g_k,\lambda_k)$ with a NGFP in the UV onto ``bare'' trajectories $(\check{g}_{\Lambda},\check{\lambda}_{\Lambda})$ which, too, possess a fixed point.
A $\Lambda$-dependent transformation would not have this property in general.
In fact, while admittedly somewhat artificial\footnote{See also the footnote in Subsection \ref{conceptual}. }, 
the choice $M=const$ realizes the possibility of having
{\it{a fixed point on the effective side, but a more complicated RG behavior on the bare side:}}
the bare parameters (even the essential ones) would keep running for $ \Lambda\rightarrow\infty$.
Nevertheless, their possibly  complicated behavior has a simple image on the effective side, 
namely a $\Gamma_k$-trajectory running into a fixed point.

\noindent
{\bf{(B) Solving for the bare parameters.}} In Fig.3 we show the result of numerically solving \eqref{qeggl}
for the bare couplings as a function of the effective ones. One finds a well defined pair
$(\check{g},\check{\lambda})$ for all $g>0$ and $\lambda<1/2$.
In the figure two values of $c$ or, equivalently of $Q$ are employed. For $Q=0$ the difference between bare and effective
parameters is small, except close to the singular boundary at $\lambda=1/2$.
The other example with $Q=-5\pi$ is typical for moderately large values of $|Q|$ where 
$(\check{g},\check{\lambda})$ differs
significantly from $(g,\lambda)$, but the map is still one-to-one. For extremely large values of $|Q|$ the one-to-one
correspondence breaks down (not shown). We shall not employ such values in the following.

\noindent
{\bf{(C) Bare fixed points.}} Next we apply the transformation $(g,\lambda)\mapsto (\check{g},\check{\lambda})$ to an ``effective'' RG trajectory. Since with $M\propto \Lambda$
the transformation has no explicit $\Lambda$-dependence , the fixed point behavior 
$\lim_{k\rightarrow\infty}(g_k,\lambda_k)=(g_*,\lambda_*)$ is mapped onto an analogous fixed 
point behavior at the bare level: 
$\lim_{\Lambda\rightarrow\infty}(\check{g}_{\Lambda},\check{\lambda}_{\Lambda})=
(\check{g}_*,\check{\lambda}_*)$. The image of the GFP is always at $\check{g}_*=\check{\lambda}_*=0$,
while the coordinates of the ``bare'' NGFP, $\check{g}_*$ and $\check{\lambda}_*$, depend on the value
of $Q$. The NGFP is an inner point of the domain in which the map 
$(g,\lambda)\mapsto (\check{g},\check{\lambda})$ is defined and is differentiable. Its Jacobian matrix
$\partial(g,\lambda)/\partial(\check{g},\check{\lambda})$ is non-singular there which entails that the
critical exponents of the NGFP are identical for the ``effective'' and the ``bare'' flow.
The same is {\it{not}} true for the GFP since it is located on the boundary of the corresponding
space of couplings (see below).

\noindent
{\bf{(D) Phase portrait of the bare flow.}} In Fig.4 we present the result of applying the map
$(g,\lambda)\mapsto (\check{g},\check{\lambda})$ to a set of representative ``effective'' 
RG trajectories on the half plane $g>0$. Fig 4(a) shows the ``effective'' flow, while the plots
(b),(c) and (d) correspond to the ``bare'' flow for three different values of Q. Since the map
does not change the critical exponents of the NGFP, the bare trajectories, too, have the typical spiral form.
However, contrary to the effective one, the bare cosmological constant at the NGFP can also
be negative or zero for suitable choices of Q. Here we emphasize again that all choices are 
physically equivalent. Varying Q simply amounts to shifting contributions back and forth
between the action and the measure.

\noindent
{\bf{(E) Vicinity of the GFP.}}  Near the GFP we may expand the relation between bare and
effective couplings. From \eqref{5.101} we obtain, in leading order,
$g = \check{g} + \mathcal{O}(\check{g}^2,\check{\lambda}^2) $, along with
\begin{equation}\label{5.104}
\lambda =\check{\lambda}+ \frac{\check{g}}{4\pi}\;\Big( Q-5\ln{\check{g}} \Big)
- \frac{\check{g}\check{\lambda}}{6\pi}\Big[ 3-Q+5\ln{\check{g}}\Big]
+ \mathcal{O}(\check{g}^2,\check{\lambda}^2) 
\end{equation}
Inverting yields the bare quantities  $\check{g} = g +\mathcal{O}(g^2,\lambda^2)$ and
\begin{equation}\label{5.105}
\check{\lambda} = \lambda 
+\frac{g}{4\pi}\;\Big(5\ln{\check{g}}-Q \Big)
+\frac{g\lambda}{6\pi}\;\Big(5\ln{\check{g}}+3-Q\Big)+\mathcal{O}(g^2,\lambda^2)
\end{equation}
These expansions are the first few terms of a power-log series. They make it explicit that
because of the logarithms of Newton's constant the relationship between $(g,\lambda)$ and
$(\check{g},\check{\lambda})$ is not analytic; it could not be found in a perturbation theory-like 
power series expansion about the GFP.

The $\Gamma_k$-trajectories linearized about the GFP are given by \cite{h3, entropy}
\begin{align}
g_k &= g_T\;\Big(\frac{k}{k_T}\Big)^2
\label{5.106}\\
\lambda_k &= \frac{1}{2}\lambda_T\;\Big[ \Big(\frac{k}{k_T}\Big)^2+\Big(\frac{k_T}{k}\Big)^2 \Big]
\label{5.106a}
\end{align}
where $g_T,\lambda_T$ and $k_T$ are constants.\footnote{The constants $g_T$ and $\lambda_T$ are the
coordinates at the ``turning point'' at which, by definition, $\beta_{\lambda}(g_T,\lambda_T)=0$.
The parameter $k_T$ is the scale at which the trajectory passes this point \cite{h3, entropy}.}
Inserting \eqref{5.106}, \eqref{5.106a}, with $k$ replaced by $\Lambda$ into the above equations we obtain the 
``bare'' trajectories $\Lambda\mapsto (\check{g}_{\Lambda},\check{\lambda}_{\Lambda})$ near the GFP:
\begin{align}\label{5.107} 
\check{g}_{\Lambda} = & g_T\Big(\frac{\Lambda}{k_T}\Big)^2
\nonumber\\
\check{\lambda}_{\Lambda} = &\frac{\lambda_T}{2}\; 
\Big[ \Big(\frac{\Lambda}{k_T}\Big)^2+\Big(\frac{k_T}{\Lambda}\Big)^2 \Big]\;
\Big\{ 1+ \frac{g_T}{6\pi}\Big(\frac{\Lambda}{k_T}\Big)^2\;\Big(     10\ln{(\Lambda/k_T)}+3+q \Big)\Big\}+
\nonumber\\
+& \frac{g_T}{4\pi}\Big(\frac{\Lambda}{k_T}\Big)^2\;\Big[  10\ln{(\Lambda/k_T)}+q \Big]
\end{align}
Here we abbreviated $q\equiv5\ln{g_T}-Q$. Note that $\check{\lambda}_{\Lambda}$ contains terms proportional
to $\ln{\Lambda}$ and $\Lambda^4\ln{\Lambda}$, that is, {\it{the RG running of the bare cosmological
and Newton constant near the GFP is not a pure power law but has logarithmic corrections.}} The
analogous running at the effective level, eqs.\eqref{5.106}, is of power law-type, however. This 
difference in behavior could not occur if the relation between bare and effective parameters was
given by a smooth map. This condition fails to be satisfied on the line $\check{g}=0$ on which
the GFP is situated.

\noindent
{\bf{To summarize:}} Both the ``effective'' and the ``bare''  NGFP are inner points of the corresponding
coupling constant space. The flow in the vicinity of one is the diffeomorphic image of the flow near the other. The RG running of the respective scaling fields is $\propto k^{-\theta}$ and 
$\propto \Lambda^{-\theta}$, respectively, with the same critical exponents $\theta$.
The ``bare'' GFP is located on the boundary of the domain on which the map from the effective to bare coupling
is defined. In its vicinity (on the half plane with $\check{g}>0$) the ``bare'' running is characterized by
logarithmically corrected power laws. The ``effective'' GFP, on the other hand, shows pure power
law scaling.

\begin{figure}
\begin{tabular}{lll}
\centering
\small{\bf{(a)}} & {}& {} \\ 
\includegraphics[width=6cm,height=4.5cm]{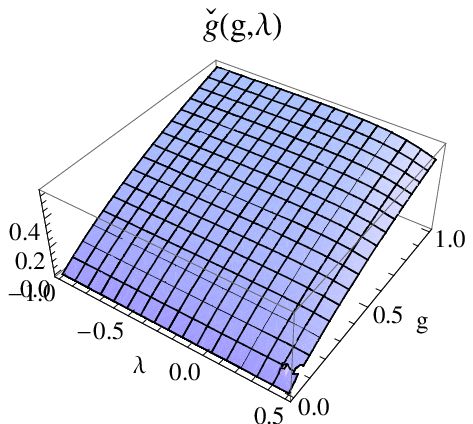} & { }& \includegraphics[width=6cm,height=4.5cm]{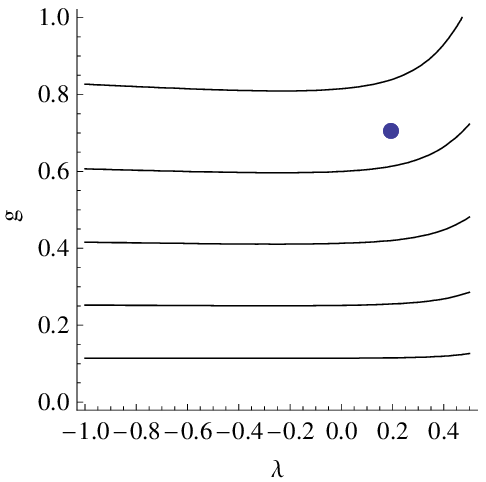} \\
\small{\bf{(b)}} &{}& {} \\
\includegraphics[width=6cm,height=4.5cm]{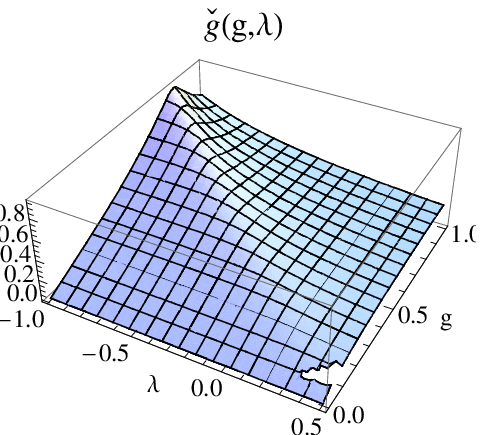} &{ }&\includegraphics[width=6cm,height=4.5cm]{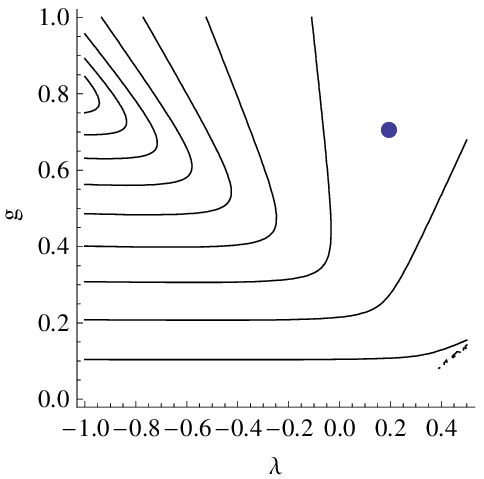}\\
\small{\bf{(c)}} & {}& {} \\ 
\includegraphics[width=6cm,height=4.5cm]{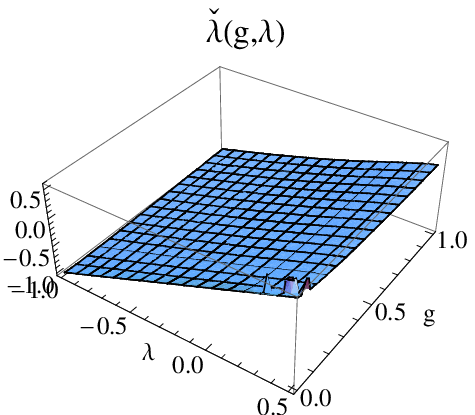} &{ }&\includegraphics[width=6cm,height=4.5cm]{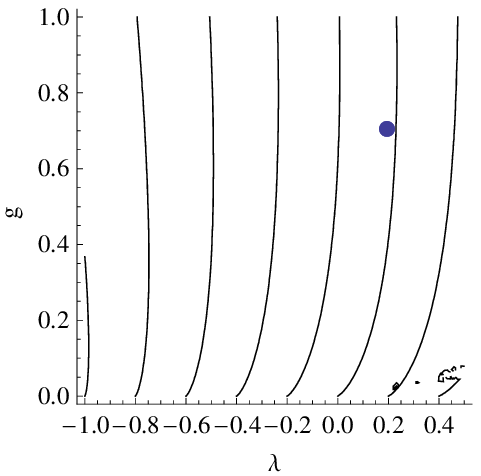}\\
\small{\bf{(d)}} &{}& {} \\
\includegraphics[width=6cm,height=4.5cm]{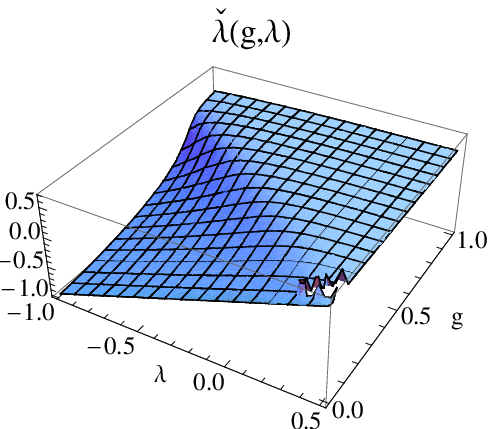} &{ }&\includegraphics[width=6cm,height=4.5cm]{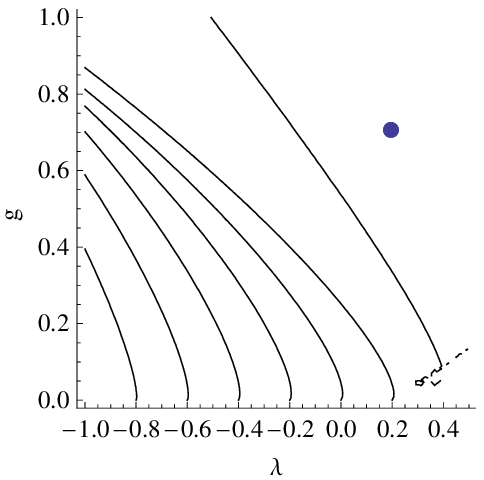}
\end{tabular}
\caption{\small{The bare parameters are shown in dependence on the effective ones, $g$ and $\lambda$, for two values of $Q$.
Fig.(a) $\check{g}$ for $Q=0$; Fig.(b) $\check{g}$ for $Q=-5\pi$; Fig.(c) $\check{\lambda}$ for $Q=0$; 
Fig.(d) $\check{\lambda}$ for $Q=-5\pi$. The results are displayed both as a 3D and a contour plot.
The blue dot in the plots of the right column marks the NGFP.}}
\end{figure}

\begin{figure}
\centering
\begin{tabular}{ll}
\bf{(a)} & \bf{(b)}\\
\includegraphics[width=7.5cm,height=5cm]{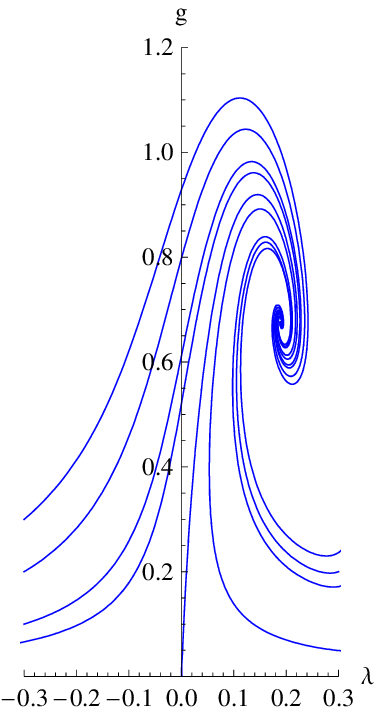} & \includegraphics[width=7.5cm,height=5cm]{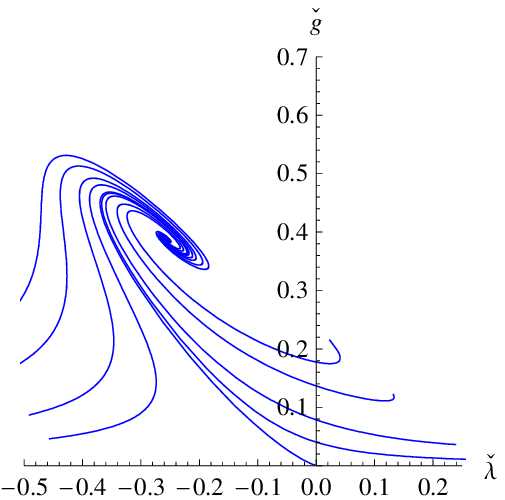}\\
 \bf{(c)}& \bf{(d)}\\
\includegraphics[width=7.5cm,height=5cm]{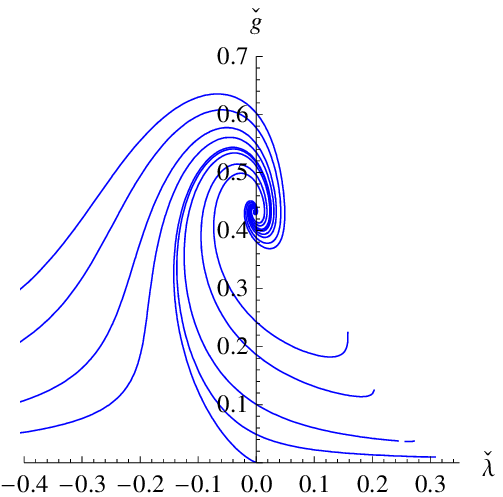} & \includegraphics[width=7.55cm,height=5cm]{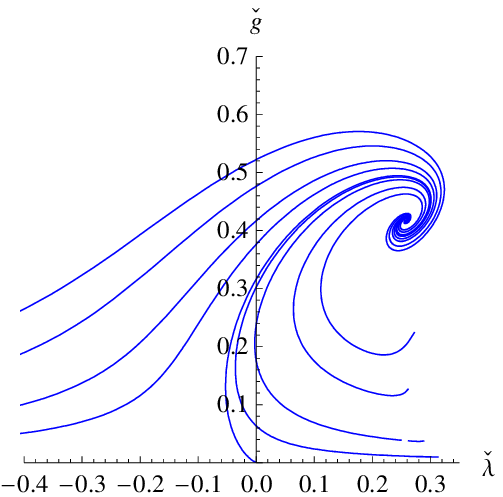}
\end{tabular}
\caption{\small{The diagram (a) shows the phase portrait of the effective RG flow on the $(g,\lambda)$-plane. The other diagrams are 
its image on the $(\check{g},\check{\lambda})$-plane of  bare parameters for three different values of $Q$, namely
(b) $Q=+1$ , (c) $Q=-0.1167$ where $\check{\lambda}_*=0$, and (d) $Q=-1$, respectively. }}
\end{figure}


\section{Summary, discussion, and outlook}\label{sectionsix}

{\bf{(A) The reconstruction problem.}} In this paper we analyzed the problem of how, being given a running
effective action, the underlying quantum system it stems from can be reconstructed.
Working in a path integral context, we showed explicitly that, after specifying a UV regularization
scheme and a measure, every solution of the flow equation for the effective average action 
(without an UV cutoff) gives rise to a regularized path integral, and to a (UV cutoff dependent)
bare action in particular.

\noindent
{\bf{(B) Completing the Asymptotic Safety program.}} While the discussion is completely general, this
work was motivated by the Asymptotic Safety program in Quantum Einstein Gravity. As to yet the 
investigations based upon the EAA focused on computing RG trajectories of the $\Lambda$-free FRGE and
establishing the existence of a non-Gaussian fixed point. The present work aims at completing the 
Asymptotic Safety program in the sense of (re)constructing the, yet unknown, quantum system which we
implicitly quantize by picking a solution of the flow equation. In fact, in our approach the primary
definition of ``QEG'' is in terms of an RG trajectory of the EAA that emanates from the fixed point.
The advantage of this strategy, defining the theory in terms of an effective rather than bare action,
is that it automatically guarantees an ``asymptotically safe'' high energy behavior. The
disadvantage is that in order to complete the Asymptotic Safety program, that is, to find the 
underlying microscopic theory, extra work is needed.

\noindent
{\bf{(C) Towards a Hamiltonian description.}} Once we know the microscopic, i.e. bare action we can 
attempt a kind of ``Legendre transformation'' to find appropriate phase space variables, a
microscopic Hamiltonian, and thus a canonical description of the bare theory. Only 
at this level we can identify the degrees of freedom that got quantized, as well as their fundamental 
interactions. Since the Hamiltonian is unlikely to turn out quadratic in the momenta, the 
``Legendre tansformation'' involved is to be understood as a generalized, i.e. quantum
mechanical one. In the simplest case it consists in reformulating a given configuration space path integral 
$\int\mathcal{D}\,\Phi \exp{\big(iS[\Phi]\big)}$ as a phase space integral
$\int\mathcal{D}\Phi\int\mathcal{D}\Pi \exp{\big(i\int\Pi\dot{\Phi}-H[\Pi,\Phi]\big)}$.
With other words, we must undo the integrating out of the momenta.

However, given the complexity of $\Gamma_*$ which most probably contains higher derivatives and non-local
terms a generalized, Ostrogradski-type phase space formalism will emerge presumably.

Being interested in a canonical description of the ``bare'' NGFP action one might wonder if there
exists an alternative formalism which deals directly with the RG flow of Hamiltonians rather
than Lagrangians.
It seems that there hardly can be a practicable approach of this kind which is  similar in spirit to the
EAA. The reason is as follows. 

If we apply a coarse graining step to an action which contains only, say, 
first derivatives of the field, the result will contain higher derivatives in general. This poses
no special problem in a Lagrangian setting, but for the Hamiltonian formalism it implies that
new momentum variables must be introduced. As a result, the coarse grained Hamiltonian
``lives'' on a different phase space (in the sense of Ostrogradski's method) than the original one.
Therefore, at least in a straightforward interpretation, there is no Hamiltonian analog of the 
flow on the space of actions. For this reason there is probably no simple way of getting around
the ``reconstruction problem''.

However, the above discussion does not contradict other approaches where the renormalization 
procedure could be applied
in a Hamiltonian description \cite{zapcor} since there the coarse graining is  performed in space 
(rather than spacetime) only. 

\noindent
{\bf{(D) Degrees of freedom vs. carrier fields.}} One should emphasize that it is by no means clear from the 
outset what kind of fundamental degrees of freedom will be found in this Hamiltonian analysis. 
In our approach the only nontrivial input is the theory space, the space of functionals on which
the renormalization group operates. Having fixed this space a FRGE can be written down, 
the resulting flow can be computed, its fixed point(s) $\Gamma_*$  can be identified, and the associated asymptotically safe
field theories can  be defined {\it{without any additional input}}.\footnote{Apart from the physically 
irrelevant details of the coarse graining scheme.} As a consequence, the only statement about
the degrees of freedom in these theories which we can make on general grounds is that they can be 
``carried'' by precisely those fields on which $\Gamma_k$ depends.
(In the case at hand, theory space contains all functionals $\Gamma[g,\bar{g},\xi,\bar{\xi}]$ which 
are invariant under diffeomorphisms.) Clearly, just knowing the carrier field but not the action, here 
$\Gamma_*$, tells us comparatively little about the degrees of freedom\footnote{Recall, for instance,
how the structure of propagating modes changes when higher derivative or nonlocal terms are added to
some conventional action.}. The action $\Gamma_*$, however, is a {\it{prediction}} of the theory,
not an input. From this point of view it is quite nontrivial that QEG was found to have  RG trajectories which
indeed describe classical General Relativity on macroscopic scales.

\noindent
{\bf{(E) This work.}} In the present paper we took a few first steps towards solving the reconstruction
problem. In particular we demonstrated that the information contained in the EAA without a UV cutoff is 
sufficient to define a regularized path integral representation of the underlying theory with
a well defined limit $\Lambda\rightarrow\infty$.

The construction requires a certain amount of EAA technology which we provided in Section \ref{eaaUV}.
In particular we explained why, using a nonsingular coarse graining operator $\mathcal{R}_k$, the EAA does not
precisely correspond to the familiar picture of a momentum shell integration, a continuum analog of the 
Kadanoff-Wilson block spin transformation. For understanding how a solution of the $\Lambda$-free 
FRGE, the corresponding solution of the FRGE with UV cutoff, and the bare action are interrelated 
it is important to appreciate this difference. In Section \ref{eaaUV} we explained the relationship between $\Gamma_k$ and $\Gamma_{k,\Lambda}$, and in 
Section \ref{reconstructing} we showed how a given $\Gamma_k$  trajectory gives rise to a trajectory
$S_{\Lambda}$ of bare actions. In Section \ref{COSMO} we illustrated the method in a technically simple
context which is physically relevant in its own right, namely the running cosmological constant induced by 
a scalar matter field. We contrasted the running effective and  bare cosmological constant and
saw, for instance, that the former is always positive at high momentum scales, while the latter can be
positive, negative, or even zero depending on the normalization of the measure. This is consistent with the 
interpretation of the effective one as the physical cosmological constant at a given scale, while its bare
counterpart is completely unphysical and may not be used for purposes of RG improvement.

Finally, in Section \ref{QEGtruncation}, we investigated  QEG in the Einstein-Hilbert truncation. 
We constructed the map relating the effective to the bare Newton  and cosmological constant, and we
analyzed the properties on the ``bare'' RG flow. We saw in particular that the ``effective'' NGFP
maps onto a corresponding ``bare'' one; in its vicinity the scaling fields show a power law running with the
same critical exponents as at the effective level. The situation is different for the GFP which is a 
boundary point of parameter space. The pure power laws of the ``effective'' flow receive logarithmic
corrections on the ``bare'' side.

\noindent
{\bf{(F) Outlook: conformally reduced gravity.}} Before closing let us mention some extensions of the work 
described here. Clearly it would be interesting to construct the bare QEG action in a more general truncation.
A first investigation in this direction \cite{elisa2} has been performed in the context of conformally 
reduced gravity \cite{creh1,creh2} in which only the conformal factor of the metric is quantized.
The simplicity of the model allows for the use of comparatively general truncations. Using the local 
potential approximation in which the $\Gamma_k$ for the conformal factor $\varphi$ is taken to  be 
of the type $\int \tfrac{1}{2}(\partial\varphi)^2+Y_k(\varphi)$ with a running potential $Y_k$ it has
been shown \cite{creh2} that there exists a NGFP on the infinite dimensional space of the $Y$'s; the
fixed point potential $Y_*$ was found to be a pure $\varphi^4$-monomial (for the ${R}^4$ topology):
$Y_*\propto \varphi^4$.

Using the method of the present paper one can now determine the corresponding bare potential function 
$\check{Y}_*$; one finds \cite{elisa2}
\begin{equation}
\check{Y}_*(\varphi)\propto \varphi^4\ln{\varphi}
\end{equation}
Remarkably, while this potential is of the familiar Coleman-Weinberg form,
it is here part of the {\it{bare}} action; it  corresponds to a simple $\varphi^4$ monomial in the
effective one . Thus, as compared to a standard scalar matter field theory, the situation is exactly
inverted.

It is not difficult to understand how this comes about: The difference $\Gamma_*-S_*$ is given by a supertrace
$\textrm{STr}[\cdots]$ which is nothing but a differentiated  one-loop determinant.
As a consequence, $\Gamma_*$ and $S_*$ differ precisely by terms typical of a one loop effective action.
For a scalar they include the potential term $\varphi^4\ln{ \varphi}$, but also nonlocal terms
(not considered here) such as $\int \varphi^2 f(-\Box)\varphi^2$, say. Hence a $\varphi^4$ term in
$\Gamma_*$ unavoidably amounts  to a Coleman-Weinberg term in $S_*$, at least within the
truncation considered.

\noindent
{\bf{(G) Outlook: Yang-Mills theory and nonlinear sigma models.}} Leaving aside gravity,
in  future work it will be interesting to analyze higher dimensional Yang-Mills theory along the same 
lines. In fact, in ref.\cite{nonabavact} the effective average action of $d$-dimensional Yang-Mills theory
was considered in a simple $\int (F^a_{\mu\nu})^2$-truncation. According to this truncation\footnote{
For a generalization see also \cite{giesymfp}.}, $\Gamma_k$ has a NGFP in the UV if $4<d<24$.
Inspired by the structure of the one-loop effective action in Yang-Mills  theory one would expect that the 
``bare'' counterpart of the $\int (F^a_{\mu\nu})^2$-fixed point should contain terms like 
$\int (F^a_{\mu\nu})^2\ln{(F^a_{\mu\nu})^2}$, and also nonlocal ones such as 
$\int F^a_{\mu\nu}f(-D^2)F^a_{\mu\nu}$. For the following reason it is of some practical 
importance to find out whether this is actually the case in a sufficiently general, reliable truncation.
It seems comparatively easy to perform Monte-Carlo simulations in $d=5$, say, so that one could
possibly get an independent confirmation of the results obtained from the average action.
However, the problem is that a priori we do not really know which bare theory should be simulated
in order to arrive at the lattice version of the average action results. The present analysis of this paper suggests
 that if  Yang-Mills theory is asymptotically safe in $d=5$, the effective  fixed point action
$\Gamma_*$ might be simple, but $S_*$ could contain ``exotic'' nonlinear and nonlocal terms. If so, it
is conceivable that $S_*$ is sufficiently different from $\int (F^a_{\mu\nu})^2$ to belong to a new
universality class. In this case a Monte-Carlo simulation based upon the conventional Wilson gauge
field action might not find a NGFP, while it should show up when a discretized version of $S_*$ is used.

Completely analogous remarks apply to the nonlinear sigma model in $d>2$ which, according to the lowest
order truncation of the EAA, is asymptotically safe too \cite{nonlinsig}.

\bigskip
\noindent
{\bf{Acknowledgments.}}
E.M. would like to thank the German Academic Exchange Service (DAAD) for support.

\newpage

\appendix
\section{Appendix}\label{appendix}
\setcounter{equation}{0}
\renewcommand{\theequation}{\ref{appendix}.\arabic{equation}}

In this appendix we evaluate the functional trace appearing in eq.(\ref{qegDE}),
\begin{equation}\label{a1}
T[\bar{g}]\equiv\frac{1}{2}\textrm{STr}_{\Lambda}\ln{\Big[\Big(\widetilde{S}_{\Lambda}^{(2)}+\widehat{\mathcal{R}}_{\Lambda}\Big)[0,0,0;\bar{g}]\;\mathcal{N}^{-1}\Big]}
\end{equation}
We perform the derivative expansion up to the second order.

Since the truncated functional $\widetilde{S}_{\Lambda}$ has the same structure as the ansatz for $\EA{k}$, we can easily obtain 
the Hessian $\widetilde{S}_{\Lambda}^{(2)}$ from $\EA{k}^{(2)}$ by replacing $G_k\rightarrow \check{G}_{\Lambda}$ and 
$\bar{\lambda}_k\rightarrow \check{\bar{\lambda}}_{\Lambda}$. The Hessian $\EA{k}^{(2)}$ has been worked out in \cite{mr}
in order to derive the FRGE so that we can take advantage of the results obtained there. As in \cite{mr}, we partly diagonalize 
the matrix  $\widetilde{S}_{\Lambda}^{(2)}$ by
$(i)$ pulling the trace out of the field $h_{\mu\nu}$, defining 
$\hat{h}_{\mu\nu}\equiv\bar{h}_{\mu\nu}-d^{-1}\bar{g}_{\mu\nu}\phi$ with $\phi\equiv\bar{g}^{\mu\nu}\bar{h}_{\mu\nu}$ and
$\bar{g}^{\mu\nu}\hat{h}_{\mu\nu}=0$, and $(ii)$ assuming the background metric to correspond  to a sphere $\mathrm{S}^d$ which,
as in \cite{mr}, means no loss of generality. As a result, the supertrace (\ref{a1}) boils down to three separate
traces over symmetric tensor fields, scalars, and vector fields, respectively. The operators 
$\widetilde{S}_{\Lambda}^{(2)}$ and $\widehat{\mathcal{R}}_{\Lambda}$ in the respective sectors can be read off from
the formulae in Section 4 of ref.\cite{mr}. In this way we obtain
%
\begin{align}\label{a2}
T[g] & = \frac{1}{2}\;\textrm{Tr}_{\Lambda}^T\ln{\Big\{ \frac{1}{32\pi\check{G}_{\Lambda}M^d}
\Big[-D^2+\Lambda^2}\;R^{(0)}\Big(-\frac{D^2}{\Lambda^2}\Big)-2\check{\bar{\lambda}}_{\Lambda}+C_TR \Big]\Big\}
\nonumber\\
{} &+ \frac{1}{2}\;\textrm{Tr}_{\Lambda}^S\ln{\Big\{\Big(\frac{d-2}{2d}\Big) \frac{1}{32\pi\check{G}_{\Lambda}M^d}
\Big[-D^2+\Lambda^2\;R^{(0)}\Big(-\frac{D^2}{\Lambda^2}\Big)-2\check{\bar{\lambda}}_{\Lambda}+C_SR \Big]\Big\}}
\nonumber\\
{} &- \textrm{Tr}_{\Lambda}^V\ln{\Big\{\frac{1}{M^2}
\Big[-D^2+\Lambda^2\;R^{(0)}\Big(-\frac{D^2}{\Lambda^2}\Big)+C_VR \Big]\Big\}}
\end{align}
Here, $\textrm{Tr}_{\Lambda}^{T,S,V}[\cdots]\equiv\textrm{Tr}^{T,S,V}[\theta(\Lambda^2+D^2)(\cdots)]$
denotes the regularized trace in the tensor, scalar, and vector sector, respectively, and the constants are defined as
in \cite{mr}:
\begin{equation}\label{a3}
C_T=\frac{d(d-3)+4}{d(d-1)},\quad C_S=\frac{d-4}{d},\quad C_V=-\frac{1}{d}
\end{equation}
The first, second, and third trace in (\ref{a2}) stems from the $\hat{h}_{\mu\nu}$, $\phi$, and the ghost fluctuations,
respectively. As the background metric was identified with a sphere of radius $r$,
the curvature scalar $R$ in (\ref{a2}) is given by $R=d(d-1)/r^2$. Our task is to expand the traces in powers of $r$, 
retaining only the terms proportional to $r^d$ and $r^{d-2}$.
They allow us to unambiguously identify the prefactors $B_0$ and $B_1$ of $\vol\sqrt{g}$ and $\vol\sqrt{g}R$, respectively.

At this point we must pick a concrete function 
$\mathcal{R}_{\Lambda}(p^2)\equiv\Lambda^2{R}^{(0)}(\tfrac{p^2}{\Lambda^2})$. We use the ``optimized"
cutoff defined in (\ref{bk0}) since $\EA{\Lambda,\Lambda}=\EA{\Lambda}$ holds true exactly then. Under the traces of (\ref{a2})
the eigenvalues of $-D^2$ are restricted to be smaller than $\Lambda^2$. Therefore (\ref{bk0}) implies that, under the traces,
$-D^2+\Lambda^2\;{R}^{(0)}(-\tfrac{D^2}{\Lambda^2})=\Lambda^2$. This leads to a considerable simplification:
\begin{eqnarray}\label{a4}
T[g] &=& \frac{1}{2}\ln{\Big\{ \frac{1}{32\pi\check{G}_{\Lambda}M^d}
\Big[\Lambda^2-2\check{\bar{\lambda}}_{\Lambda}+C_TR \Big]\Big\}}\;\textrm{Tr}_T[\theta(\Lambda^2+D^2)]
\nonumber\\
&&{}+\frac{1}{2}\ln{\Big\{\Big(\frac{d-2}{2d}\Big) \frac{1}{32\pi\check{G}_{\Lambda}M^d}
\Big[\Lambda^2-2\check{\bar{\lambda}}_{\Lambda}+C_SR \Big]\Big\}}\;\textrm{Tr}_S[\theta(\Lambda^2+D^2)]
\nonumber\\
&&{}-\ln{\Big\{\frac{1}{M^2}
\Big[\Lambda^2+C_VR \Big]\Big\}}\;\textrm{Tr}_V[\theta(\Lambda^2+D^2)]
\end{eqnarray}
The derivative expansion of the traces involving the step function can be found with standard heat kernel techniques.
Up to second derivatives of the metric it reads\footnote{This formula is most easily obtained by making use of eq.(4.27)
in ref.\cite{mr} for the special case $W(z)=\theta(\Lambda^2-z)$.}
\begin{equation}\label{a5}
\begin{split}
\textrm{Tr}_{T,S,V}[\theta(\Lambda^2+D^2)] & = \Big(\frac{1}{4\pi}\Big)^{d/2}\frac{1}{\Gamma(\frac{d}{2}+1)}\;
\textrm{tr}(I_{T,S,V})\;\Big\{\Lambda^d\vol\sqrt{g}+\\
& \quad +\frac{d}{12}\Lambda^{d-2}\vol\sqrt{g}R +O(R^2)     \Big\}
\end{split}
\end{equation}
Here the algebraic trace $\textrm{tr}(I_{T,S,V})$ counts the number of independent field components in the $T,S$, and $V$ sectors; it
equals $\textrm{tr}(I_T)=(d-1)(d+2)/2$, $\textrm{tr}(I_S)=1$, and $\textrm{tr}(I_V)=d$, respectively.

Now it is a matter of straightforward algebra to insert (\ref{a5}) and (\ref{a3}) into (\ref{a4}), and to expand in powers 
of $r$, noting that $\sqrt{g}\propto r^d$ and $R\propto r^{-2}$. The final result for the supertrace is then indeed found to have
the same structure as the RHS of (\ref{qegDE}) so that we can read off the coefficients $B_0$ and $B_1$:
\begin{subequations}\label{a6}
\begin{align}
B_0 &= \frac{1}{2(4\pi)^{d/2}\Gamma(d/2)}\Big[(d+1)\ln{(1-2\check{\lambda}_{\Lambda})}-(d+1)\ln{(\check{g}_{\Lambda})} +
Q_{\Lambda}\Big]\label{abb}\\
B_1 &= \frac{d}{12}B_0 +\Delta B_1\label{ab1}\\
\Delta B_1 &\equiv 
\Big(\frac{1}{4\pi}\Big)^{d/2}\;\;
\frac{d(d-1)+4(1-2\check{\lambda}_{\Lambda})}{2d\;\Gamma(\frac{d}{2})\;(1-2\check{\lambda}_{\Lambda})}
\label{adb}\\ 
Q_{\Lambda}& \equiv \big[d(d+1)-8\big]\ln{(\Lambda/M)}+b_0
\label{aqq}
\end{align}
\end{subequations}
Here $b_0\equiv -(d+1)\ln{(32\pi)}+2d^{-1}\ln{\big((d-2)/(2d)\big)}$. For  $d=4$ the above results reduce to
those in the eqs.(\ref{Bs}) of the main text.

\end{document}